\documentclass[a4paper,11pt,showpacs,amsmath,nofootinbib,superscriptaddress]{revtex4}
\usepackage{amssymb}
\usepackage[hypertex]{hyperref}
\usepackage{graphicx}
\usepackage{epstopdf}
\usepackage{bm}
\usepackage{epsfig}
\usepackage{graphics}
\usepackage{xspace}
\usepackage{amsfonts}
\usepackage{comment}

\def\OMIT#1{}

\newcommand{\nn}{\nonumber}

\newcommand{\bea}{\begin{eqnarray}}
\newcommand{\eea}{\end{eqnarray}}

\newcommand{\mcdot}{\!\cdot\!}

\newcommand{\half}{\frac{1}{2}}

\newcommand{\muj}{\mu_j}

\newcommand{\cO}{{\cal O}}

\newcommand{\la}{\langle}
\newcommand{\ra}{\rangle}

\begin{document}


\title{Renormalization-group improved predictions for Higgs boson production at large $p_T$}

\vspace*{1cm}

\author{Fa Peng Huang}
\affiliation{School of Physics and State Key
Laboratory of Nuclear Physics and Technology, Peking
University, Beijing, 100871, China}

\author{Chong Sheng Li\footnote{Electronic
address: csli@pku.edu.cn}}
\affiliation{School of Physics and State Key
Laboratory of Nuclear Physics and Technology, Peking
University, Beijing, 100871, China}
\affiliation{Center for High Energy Physics, Peking
University, Beijing, 100871, China}

\author{Hai Tao Li}
\affiliation{School of Physics and State Key
Laboratory of Nuclear Physics and Technology, Peking
University, Beijing, 100871, China}

\author{Jian Wang}
\affiliation{PRISMA Cluster of Excellence $\&$ Mainz Institut for Theoretical Physics,
Johannes Gutenberg University, D-55099 Mainz, Germany}


\begin{abstract}
 \vspace*{0.3cm}
We study the next-to-next-to-leading logarithmic order resummation  for the large $p_T$ Higgs boson production at the LHC
in the framework of soft-collinear effective theory.
We find that the  resummation effects reduce the  scale uncertainty significantly and
decrease the   QCD NLO  results by about $11\%$ in the large $p_T$ region.
The finite top quark mass effects and the effects of the NNLO singular terms  are also discussed.
\end{abstract}
\maketitle
\newpage



\section{Introduction}
\label{sec:1}
In the standard model (SM), the Higgs boson is predicted by the Higgs mechanism in which the would-be Goldstones  become the longitudinal components of the \emph{W } and \emph{Z} bosons.
Although the existence of Higgs boson has been proposed for a long time, searching for this  particle in the experiments has failed until the recent discovery at the LHC \cite{Chatrchyan:2012ufa,Aad:2012tfa}.
In general, the Higgs boson may not be responsible for the mass origin of the fermions,
and the current experimental data still allow the couplings of the Higgs boson
with the fermions to deviate from the SM ones, especially, the coupling of the Higgs boson to the top quark \cite{Chatrchyan:2013yea}.
Therefore, precise measurements of the couplings of the Higgs boson with other SM particles
will test the  Higgs mechanism in the SM  \cite{Harlander:2013oja,Azatov:2013xha,Banfi:2013yoa,Grojean:2013nya}.

The global fit method with current experiment data about the Higgs boson production and decay in various channels only provides
indirect information on the top quark Yukawa coupling, which suffers from ambiguities from unknown new particles propagating in the loops.
The most direct process to determine the coupling of the Higgs boson to the top quark is
the $t\bar{t}$ associated production $pp \to t\bar{t}H$  and single top associated production $pp \to tjH$.
However, the current abilities to measure the coupling of Higgs boson to the top quark through  $t\bar{t}$ associated production are still weak \cite{Chatrchyan:2013yea,ATLAS-CONF-2014-011} because of its small production cross section and complicated final states with copious decay products.
The single top associated production has even smaller cross section because of the electro-weak interactions there, and
is very challenging to measure.

Recently, a complementary  method to determine the coupling of the Higgs boson to the top quark has
been proposed in Refs. \cite{Azatov:2013xha, Banfi:2013yoa,Grojean:2013nya} by investigating  the
large $p_T$ behavior of the Higgs boson in the process $pp \to H+X$ with $ H \to ZZ^{*} \to l^+l^-l^+l^- $.
This method is feasible because the top quark mass can not be taken to be infinity when the Higgs boson has a large $p_T$.
The top quark Yukawa coupling can be detected from the measurement of the variable \cite{Azatov:2013xha}
\begin{equation}\label{eq:r}
    r_{\pm}= \frac{N^+/N^-}{\sigma_{\rm SM}^+/\sigma_{\rm SM}^-},
\end{equation}
where $N^{\pm}$ is the number of events in which the Higgs boson $p_T$ is larger or smaller than a critical value $P_T$, for example, $P_T=$300 GeV.
$\sigma_{\rm SM}^{\pm }$ is the corresponding theoretical predictions in the SM.
It is pointed out that  \cite{Azatov:2013xha} the $K$ factor, defined as the ratio of higher order results to the LO ones,
for the Higgs boson $p_T$ distribution is roughly $p_T$ independent and very large, about 2,
and that the resummation effects are negligible in the $p_T$ range they considered.
All these arguments are based on the calculation by the HqT program \cite{deFlorian:2011xf}.
However, the resummation scheme used in the HqT program is only valid in the small $p_T$ region, which is much less than 100 GeV.
The resummation prediction on the Higgs boson $p_T$ distribution in the large $p_T$ region,  larger than 100 GeV,
is investigated using the traditional method at next-to-leading-logarithm (NLL) \cite{Florian:2005rr}.
The resumed logarithms are different in the small and large $p_T$ regions.
When the Higgs boson $p_T$ is small, the threshold region is defined as $z=M_H^2/s \to 1$, where $\sqrt{s}$ is the center-of-mass energy
of the colliding partons. In the threshold region, only the soft gluon radiation is allowed, which leads to large logarithms $\alpha_s^n \ln^{2n-m} (1-z)$.
In contrast, in the large $p_T$ regions, the large logarithms are $\alpha_s^n \ln^{2n-m} (1-y)$ with $y=(p_T+m_T)^2/s$,
where $m_T=\sqrt{p_T^2+M_H^2}$ \cite{Florian:2005rr}.
It is easy to observe that $y \to 1$ does not necessarily lead to $z \to 1$, which means that the HqT program can not resum the
large logarithms in the large $p_T$ regions.

Notice that when the recoiling hardest jet against the Higgs boson is observed and additional jets are vetoed,
there is a new kind of large Sudakov logarithms  $\alpha_s^n \ln^{2n-m} p_T/p_T^{\rm veto}$, which can be resummed \cite{Liu:2012sz,Liu:2013hba,Boughezal:2013oha}.
If the mass of the jet is also measured, denoted as $m_J$, additional logarithms $\ln^n m_J^2/p_J^2$
have been resummed up to next-to-next-to-leading-logarithm (NNLL) \cite{Jouttenus:2013hs}.

In this paper, we provide the resummed prediction for $pp \to H+X$ at large $p_T$ regions up to NNLL, without explicit observation of a jet, in contract with the case in \cite{Jouttenus:2013hs}.
We will work in the soft-collinear effective theory (SCET)~\cite{Bauer:2000ew,Bauer:2000yr,Bauer:2001ct,Bauer:2001yt,Becher:2006nr}.
In the threshold limits  of large $p_T$ Higgs boson production,
the final state radiations and beam remnants are
highly suppressed  which  leads to final states  consisting of a Higgs boson
and an inclusive jet, as well as the remaining soft radiations,
and therefore  to the appearance of  the large logarithms in the cross section.
Then the resummation  effects   should be taken into account to obtain more precise
predictions.
The preliminarily theoretical NNLO analyses have been perfromed in Ref. \cite{Becher:2013vva}.
The resummation formalism in SCET is different from that used in Ref. \cite{Florian:2005rr}.
In the threshold region, the partonic cross section can be factorized to a hard function times a convolution between jet and soft functions.
Each part has a explicit theoretical field definition which can be calculated perturbatively.
In particular, each function contains only a single energy scale so that there is no potential large logarithms in each of them.
The relative scale hierarchy between different functions is alleviated by running from one to the other via renormalization group equations.
As a consequence, the large logarithms of the ratio of the different scales can be resummed to all orders.


In principle, the top quark mass should be kept to be finite in all the theoretical predictions in the large $p_T$ regions
\cite{Ravindran:2002dc, Harlander:2012hf,Azatov:2013xha, Grojean:2013nya}.
But because of the difficulty in calculating massive loops, this is achieved only for the LO result \cite{Ellis:1987xu,Baur:1989cm} and the NLO total cross section  expanded  in $M_H/m_t$\cite{Harlander:2009mq}.
The differential cross section is calculated only in the large top quark limit up to NLO \cite{Schmidt:1997wr,Ravindran:2002dc,deFlorian:1999zd,Glosser:2002gm,Ravindran:2002dc}.
More recently, a big progress is made by computing the NNLO total cross section of the sub-process $gg\to H+j$ \cite{Boughezal:2013uia}.
Therefore, an approximated differential cross section with finite top quark mass beyond the LO is usually used,
which is obtained by multiplying the LO differential cross section with finite top quark mass with a differential $K$ factor,
as done in Ref. \cite{Grazzini:2013mca}.
We will take into account the finite top quark mass effects in the resummation predictions following this method.

The precision prediction on  Higgs boson production at large $p_T$ regions can not only test the top quark Yukawa coupling discussed above,
but  also be  a probe  of the new physics.
For example, in the SM the large transverse momentum spectrum of the Higgs boson  produced in gluon fusion  can be quite different  from one of the minimal supersymmetric standard model \cite{Langenegger:2006wu,Bagnaschi:2011tu}.
Light particles beyond the SM can be probed  via the ratio of the partially integrated Higgs transverse momentum distribution
to the inclusive rate \cite{Arnesen:2008fb}.

This paper is organized as follows.
In Sec.~\ref{sec:kine}, we analyze the kinematics of the Higgs boson and one jet associated production  and give the definition of the threshold region.
In Sec.~\ref{sec:form}, we present the factorization and resummation formalism
in momentum space using SCET.
In Sec.~\ref{sec:nlo}, we present the hard function, jet function and soft functions at NLO. Then, we study the scale independence of the final result analytically.
In Sec.~\ref{sec:nume}, we discuss the numerical results for this process at the LHC.
We conclude in Sec.~\ref{sec:conc}.

\section{Analysis of kinematics}
\label{sec:kine}

\begin{figure}[h]
  \includegraphics[width=0.6\linewidth]{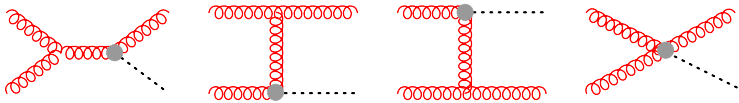}\\
  \caption{LO Feynman diagrams for the gg channel.}
  \label{fig-born}
\end{figure}

First of all, we introduce the relevant kinematical
variables needed in our analysis. The dominant partonic processes for
the Higgs boson and one jet  production are $gg \rightarrow gH$,
$gq \rightarrow gH$ and $g\bar{q}\rightarrow \bar{q}H$.
The LO Feynman diagrams for the $gg \rightarrow gH$ process are shown in Fig. \ref{fig-born}.

It is convenient to  define two
lightlike
vectors along the beam directions, $n_a$ and $n_b$, which
are
related by $n_a=\bar{n}_b$. Then, we introduce initial
collinear
fields along $n_a$ and $n_b$ to describe the collinear
particles in
the beam directions. In the center-of-mass (c.m.) frame  of the
hadronic
collision, the momenta of the incoming hadrons are given by
\begin{equation}
 P^\mu_a=E_{\rm c.m.}\frac{n^\mu_a}{2},\qquad P^\mu_b=E_{\rm
c.m.}\frac{n^\mu_b}{2}.
\end{equation}
Here $E_{\rm c.m.}$ is the c.m. energy of the
collider and we have neglected the masses of the hadrons.
The
momenta of the incoming partons, with a light-cone
momentum fraction of the hadronic momentum, are
\begin{equation}
 \tilde{p}_a = x_a E_{\rm c.m.}\frac{n^\mu_a}{2},\qquad
\tilde{p}_b= x_b E_{\rm
c.m.}\frac{n^\mu_b}{2}.
\end{equation}

At the hadronic and partonic level, the momentum conservation
gives
\begin{equation}
 P_a + P_b = q + P_X,
\end{equation}
and
\begin{equation}
 \tilde{p}_a+\tilde{p}_b=q + p_X,
\end{equation}
respectively, where $q$ is the momentum of the Higgs boson. We
define the partonic jet with jet momentum $p_X$ to be the set of
all final state partons except the Higgs boson in the partonic processes,
while the hadronic jet with jet momentum $P_X$ contains all the hadrons as
well as the beam remnants in the final state except the Higgs boson.

We also define the Mandelstam variables as
\begin{equation}
 s=(P_a+P_b)^2,\quad u=(P_a-q)^2, \quad t=(P_b-q)^2
\end{equation}
for hadrons, and
\begin{equation}
  \hat{s}=(\tilde{p}_a+\tilde{p}_b)^2,\quad
\hat{u}=(\tilde{p}_a-q)^2, \quad \hat{t}=(\tilde{p}_b-q)^2
\end{equation}
for partons, respectively. In terms of the Mandelstam
variables, the
hadronic and partonic threshold variables are defined as
\begin{eqnarray}
S_4 &\equiv& P^2_X = s+t+u-M^2_H,  \\
s_4 &\equiv& p^2_X =\hat{s}+\hat{t}+\hat{u}- M^2_H,
\end{eqnarray}
where $M_H$ is the mass of  the Higgs boson. The hadronic
threshold limit is defined as $S_4 \to 0$~\cite{Laenen:1998qw}. In
this limit, the final state radiations and beam remnants are
highly
suppressed, which leads to final states consisting of a Higgs boson
and an energetic jet, as well as the remaining soft radiations. Taking
this
limit requires $x_a\to 1,\;x_b\to 1,\; s_4 \to 0$
simultaneously,
and we get
\begin{eqnarray}\label{eqs:s4}
  S_4 &=& s_4 + \hat{s}(\frac{1}{x_ax_b}-1)+(\hat{t}-M^2_H)(\frac{1}{x_b}-1)+(\hat{u}-M^2_H)(\frac{1}{x_a}-1) \nn\\
   &\approx& s_4+\hat{s}(\bar{x}_a+\bar{x}_b)+(\hat{t}-M^2_H)\bar{x}_b+(\hat{u}-M^2_H)\bar{x}_a \nn\\
   &\approx& s_4+(-\hat{t})\bar{x}_a+(-\hat{u})\bar{x}_b,
\end{eqnarray}
where $\bar{x}_{a,b}=1-x_{a,b}$.
This expression can help to check the factorization scale invariance, which is shown in detail below.
Near the partonic  threshold, the boson must be recoiling against a jet and
there is only phase space for the jet to be nearly massless.
In this case, $p_X=p_1 + k$, where $p_1$ is the momentum of the final
state collinear partons forming the jet and $k$ is the momentum of
the soft radiations.

We note that in both hadronic and partonic threshold limit, the Higgs boson is not forced to be produced at rest, $i.e.$ it can have a large momentum.
Actually, as the momentum of the Higgs boson becomes larger and larger, the final-state phase space lies more close to the threshold limit.
We point that the definition of the partonic threshold limit $s_4/\hat{s} \to 0$ is different from the case of $y \to 1$ \cite{Florian:2005rr},
as discussed in the introduction.
They are equivalent to each other only if the momentum component $p_z$ of the Higgs boson in the partonic c.m. frame vanishes.

For  convenience, we can also write the threshold variable as
\begin{equation}
\label{eqs:s4a}
 s_4=p^2_X=(\tilde{p}_a+\tilde{p}_b-q)^2=p^2_1+2k^+
E_1+\cO(k^2),
\end{equation}
where $k^+=n_1\mcdot k$, $k$ is the momentum of soft radiations,
$E_1$ is the energy of the jet and
$n_1$ is the lightlike vector associated with the jet
direction.
In the
threshold limit~($s_4\to 0$), incomplete cancelation of the divergences between
real
and virtual corrections leads to singular distributions
$\alpha^n_s
[\ln^m(s_4/M^2_H)/s_4]_+$, with $m \leq 2n-1$. It is the
purpose of
threshold resummation to sum up these contributions to all
orders in $\alpha_s$.

The  total cross section
is given by
\begin{eqnarray}\label{eqs:main}
  \sigma &=& \int dx_a \int dx_b \int d\hat{t} \int d\hat{u} f_{i/P_a}(\mu_F,x_a)f_{j/P_b}(\mu_F,x_b)
  \frac{1}{2\hat{s}}\frac{d\hat{\sigma}_{ij}}{d\hat{t}d\hat{u}} \nn\\
   &=&\int_{0}^{p^2_{T,max}} dp_T^2 \int_{-y_{max}}^{y_{max}} dy \int_{x_{b,min}}^1 dx_b \int_0^{s_4^{max}} ds_4
   \frac{1}{2(x_bs+u-M_H^2)} f_{i/P_a}(\mu_F,x_a)f_{j/P_b}(\mu_F,x_b)
   \frac{d\hat{\sigma}_{ij}}{d\hat{t}d\hat{u}},\nn\\
\end{eqnarray}
where we have changed the integration variables into the
Higgs boson transverse momentum squared $p^2_T$, rapidity $y$, $x_b$ and $s_4$.
The regions of the integration variables are given by
\begin{eqnarray}
  p^2_{T,{\rm max}} &=& \frac{(s-M_H^2)^2}{4s}, \nn\\
  y_{\rm max} &=& \frac{1}{2}{\rm ln} \frac{1+\sqrt{1-\xi}}{1-\sqrt{1-\xi}}, \nn \\
  x_{b,{\rm min}} &=& \frac{-u}{s+t-M_H^2}, \nn\\
  s_4^{\rm max} &=& x_b(s+t-M_H^2)+u,
\end{eqnarray}
with
\begin{eqnarray}
  \xi &=&\frac{4s(p_T^2+M_H^2)}{(s+M_H^2)^2}, \nn \\
  t &=& M_H^2 -\sqrt{s}\sqrt{p_T^2+M_H^2}e^{y}, \nn \\
  u &=& M_H^2 -\sqrt{s}\sqrt{p_T^2+M_H^2}e^{-y}.
\end{eqnarray}
The other kinematical variables can be expressed in terms of these four integration variables.

\section{Factorization and Resummation Formalism in SCET}
\label{sec:form}
In the frame of SCET, we define a small expanded parameter $\lambda=\sqrt{s_4}/Q$ ($\lambda <<1$) in the threshold limit $s_4 \to 0$. Here,
Q is the characteristic energy of the hard scattering process.  The momentum of a collinear particle scales as
\begin{equation}
\mbox{collinear}: \quad p_{\mathrm{c}}^\mu  \sim  Q(1,\lambda^2,\lambda),
\end{equation}
and the  momentum of a soft particle scales as
\begin{equation}
\mbox{soft}: \quad p_{\mathrm{s}}^\mu  \sim  Q(\lambda^2,\lambda^2,\lambda^2).
\end{equation}
The soft fields scale as $\psi_{\mathrm{s}}  \sim  \lambda^3$, $A_{\mathrm{s}}  \sim  \lambda^2$, and the collinear
fermion field  $\psi_{\mathrm{c}}$ scales as $\lambda$.
The light-cone components of the collinear gluon field $A_{\mathrm{c}}^\mu$ scale the same way as
its momentum $p_{\mathrm{c}}^\mu$  in covariant gauge.

The soft gluon field are multipole expanded around $x_-$ to maintain
a consistent power counting in $\lambda$.  Thus, the soft gluon operator
depends only on  $x_-^\mu = ({\bar{n}_J}\cdot x )\frac{ { n_J^\mu}}{2}$ at leading power,
and its Fourier transform  only depends on $k_+ = { n_J} \cdot k$.
It is needed to mention that $p_s^2 \sim Q^2 \lambda^2$  is of order of the jet mass and
is assumed to be in the perturbative region.

In the limit of the infinite top quark mass, the effective Lagrangian of Higgs boson production
via gluon fusion can be written as \cite{Schmidt:1997wr}
\begin{equation}\label{eft}
\textit{L}=C_t(m_t,\mu)\frac{\alpha_s(\mu)}{12\pi }\frac{H}{\textit{v}}Tr(G_{\mu\nu}G^{\mu\nu}),
\end{equation}
with
\begin{equation}\label{ct}
  C_t=1+\frac{11\alpha_s(\mu)}{4\pi},
\end{equation}
where $C_t$ is the   Wilson coefficient at $\alpha_s$ order.
The leading power effective operator of $gg \rightarrow gH$ in SCET is  as follows,
\begin{align}\label{operator}
	\mathcal{O}_{abc}^{\alpha\beta\gamma} (x; t_1,t_2,t_J) &=
	{\cal A}_{a\perp}^{\alpha 1}(x+ t_1 \bar{n}_1) \,{\cal A}_{b\perp}^{\beta 2}(x+ t_2 \bar{n}_2) \,
	{\cal A}_{c\perp}^{\gamma J}(x+ t_J \bar{n}_J), \,
\end{align}
where ${\cal A}_{a\perp}^{\alpha,i}$ is the effective gluon field in the frame of SCET.
The corresponding hadronic operator can be  written as
\begin{equation}\label{jcurrent}
	\mathcal{J}(x) =  \int d t_1\, d t_2 \, d t_J \,
	C^{abc}_{\alpha\beta\gamma}(t_1,t_2,t_J)\, \mathcal{O}_{abc}^{\alpha\beta\gamma}(x; t_1,t_2,t_J) \,.
\end{equation}
The generic expression of  the cross section is
\begin{equation} \label{jiemian}
\mathrm{d}\sigma =\frac{1}{2s}(\frac{\alpha_s C_t}{12 \pi \textit{v} })^2  \frac{d^3 q}{(2\pi)^3 2 E_H} \sum_X (2\pi)^4 \delta^{(4)}(P_1+P_2-p_X-q)  \big |\langle  X \, | \mathcal{J}(0)  | N_1(P_1) N_2(P_2) \rangle \big |^2 .
\end{equation}
Substituting Eqs. (\ref{operator})$-$ (\ref{jcurrent}) into Eq. (\ref{jiemian}) and performing Fourier transformation, we get
\begin{multline}\label{matrix}
\mathrm{d}\sigma =\frac{1}{2s}(\frac{\alpha_s C_t}{12 \pi \textit{v} })^2  \frac{d^3 q}{(2\pi)^3 2 E_H} \,
\sum_X {\widetilde C}^{abc*}_{\alpha\beta\gamma} {\widetilde C}_{\mu\nu\rho}^{def} \\
\times  \int \mathrm{d}^4 x\, e^{-i (q x)}\,
\langle  N_1(P_1)\, N_2(P_2) |\,  \mathcal{O}_{\alpha\beta\gamma}^{abc\dagger}(x)\, |X\rangle\langle X|\, \mathcal{O}_{\mu\nu\rho}^{def}(0)  \,|  N_1(P_1)\, N_2(P_2) \rangle \,.
\end{multline}
After redefining the field to decouple the soft interactions, the operator factorizes into a collinear and a soft part
\begin{equation}\label{decouple}
\mathcal{O}_{abc}^{\alpha\beta\gamma}=\mathcal{O}^{S}\mathcal{O}_{abc}^{\alpha\beta\gamma C},
\end{equation}
where the collinear part $\mathcal{O}_{abc}^{\alpha\beta\gamma C}$ has the same form as $\mathcal{O}_{abc}^{\alpha\beta\gamma}$
in Eq. (\ref{operator}) with the collinear fields replaced by those not interacting with soft gluons,
and  the soft part $\mathcal{O}^{s}=Y_1Y_2Y_J$.  $Y_i$ is the soft Wilson lines defined as
\begin{equation}
Y_i(x) =
 \textbf{P} \exp  \left( ig \int_{-\infty}^0 dt n_i \cdot A_s^a(x+t n_i)\textbf{T}^a_i \right) ,
\end{equation}
where $\rm{\textbf{P}}$ indicates path ordering.
From here, we omit the color index for simplicity, and
we rewrite the squared amplitude in Eq. (\ref{matrix}) as
\begin{multline}\label{huang}
\langle  N_1(P_1)\, N_2(P_2) |\,  \mathcal{O}_{\alpha\beta\gamma}^{\dagger}(x)\, |X\rangle\langle X|\, \mathcal{O}_{\mu\nu\rho}(0)  \,|  N_1(P_1)\, N_2(P_2) \rangle \,
=\\
\left\langle  N_1(P_1) \left| {\cal A}_{1\alpha}^{\perp} (x) {\cal A}_{1\mu}^{\perp}(0) \right|  N_1(P_1) \right\rangle \: \times \:
\left\langle  N_2(P_2) \left| {\cal A}_{2\beta}^{\perp} (x)  {\cal A}_{2\nu}^{\perp}(0) \right|  N_2(P_2) \right\rangle \\
\times \sum_{X_c}
\langle  0 |  {{\cal A}_J^\perp}_{\gamma}(x)  | X_c \rangle
\langle  X_c |  {{\cal A}_J^\perp}_{\rho}(0) |0 \rangle
\times
\sum_{X_s}
\langle 0 | \mathcal{O}^{s\dagger}_{gg}(x)|X_s \rangle
\langle  X_s|   \mathcal{O}^{s\vphantom{\dagger}}_{gg}(0)|0 \rangle \, .
\end{multline}

Substituting the  definition of the gluon jet function, soft function, parton distribution functions (PDFs) and hard function
into Eq. (\ref{huang}),
\begin{equation}\label{jetdef}
\sum_{X_c}
\langle  0 |  {{\cal A}_J^\perp}_{\gamma}(x)  | X_c \rangle
\langle  X_c |  {{\cal A}_J^\perp}_{\rho}(0) |0 \rangle \propto (-g^\perp_{\gamma\rho})\, \int \frac{\mathrm{d}^4 p}{(2\pi)^3}\, \theta(p^0)\,  J_g(p^2) \,e^{-i\, x\, p},
\end{equation}
\begin{equation}\label{softdef}
\sum_{X_s}
\langle 0 | \mathcal{O}^{s\dagger}_{gg}(x)|X_s \rangle
\langle  X_s|   \mathcal{O}^{s\vphantom{\dagger}}_{gg}(0)|0 \rangle \propto \int_0^{\infty} \mathrm{d} k_+ \, e^{-i k_+ ( { \bar{n}_J} \cdot x)/2}\, \mathcal{S}_{gg}(k_+),
\end{equation}
\begin{equation}\label{pdfdef}
\langle  N_i(P_i)\, |\,\, (-g_{\mu\nu})\,\, {\cal A}_{i \perp}^\mu\left(
n_i \cdot x \frac{ { \bar{n}_i^\mu} }{2}\right) {\cal A}_{i \perp}^\nu(0) \,|  N_i(P_i) \rangle
=\int_{-1}^1 \frac{\mathrm{d} \xi}{\xi} f_{g/N_i}(\xi) e^{i \xi (n_i\cdot x) (\bar{n}_i \cdot P_i)/2},
\end{equation}
we obtain (up to power corrections) \cite{Beneke:2009rj,Becher:2009th}
\begin{eqnarray}\label{eqs:facmain0}
\sigma &=&\int dx_a dx_b d\hat{t}d\hat{u} \frac{1}{2\hat{s}}
f_{i/P_a}(x_a,\mu) f_{j/P_b}(x_b,\mu)
\frac{d\hat{\sigma}_{ij}^{\rm thres}}{d\hat{t}d\hat{u}}, \\
\label{eqs:facmain}
\frac{d\hat{\sigma}_{ij}^{\rm thres}}{d\hat{t}d\hat{u}} &=&
 \frac{1}{8\pi}\frac{1}{\hat{s}}
 \lambda_{0,ij} H_{ij}(\mu) \nn \\
&& \times  \int dk^+\int dp_1^2 \, \mathcal{S}(k^+,\mu) J(p_1^2,\mu)\delta(s_4-p_1^2-2k^+E_1),
\end{eqnarray}
with
\begin{eqnarray}
 \lambda_{0,gg} &=&\frac{1}{2^2 (N^2_c-1)^2} \frac{\alpha_s^3}{9 \pi  v^2} \frac{4 N_c \left(N_c^2-1\right) \left(M_H^8+{\hat{s}}^4+{\hat{t}}^4+{\hat{u}}^4\right)}{\hat{s} \hat{t} \hat{u}},\\
 \lambda_{0,gq} &=&\frac{1}{2^2 N_c (N^2_c-1)} \frac{\alpha_s^3}{9 \pi  v^2} \frac{2 \left(N_c^2-1\right) \left({\hat{s}}^2+{\hat{u}}^2\right)}{-\hat{t}},
\end{eqnarray}
where  $\lambda_{0,ij}$ is the squared  amplitude  at LO after averaging  the spins and colors.

The other channels follow the same approach to obtain the factorization formulas.
By crossing symmetry, the LO cross sections in other channels  are obtained by
\begin{eqnarray}
  \lambda_{0,g\bar{q}} &=&  \lambda_{0,gq}  (\hat{s}\leftrightarrow \hat{u}), \\
  \lambda_{0,q\bar{q}} &=&  -\lambda_{0,gq} (\hat{s}\leftrightarrow \hat{t}).
\end{eqnarray}

Here, we point out that the factorization form given in Eq.(\ref{eqs:facmain}) is only valid in the threshold limit defined by $s_4\to 0$, which means that the Higgs boson should have a large $p_T$. The traditional transverse momentum dependent factorization and resummation \cite{Collins:1984kg,Collins:2011zzd} is important when the total transverse momentum of  the Higgs boson and the recoiling jet is small, that is obvious not the same threshold region as the case we have discussed in this paper. An application of the transverse momentum resummation  in Higgs plus one jet production has been discussed in  \cite{Liu:2012sz,Liu:2013hba,Boughezal:2013oha,Sun:2014lna}.


\section{The Hard, Jet and Soft Functions at NLO }
\label{sec:nlo}
The hard, jet and soft functions describe interactions at different scales, which
can be calculated  order by order  in QCD, respectively.
At the NNLL accuracy, we need the explicit expressions of the hard, jet and soft functions up to NLO.
In this section, we summarize  the relevant analytic results of them.

\subsection{Hard functions}
\label{subsec:41}
\begin{figure}
  \includegraphics[width=0.5\linewidth]{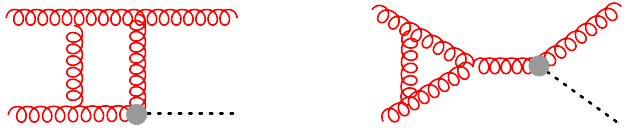}\\
  \caption{The sample one-loop Feynman diagrams for the subprocess $gg\to gH$.}
  \label{fig:virtual}
\end{figure}

The hard functions are absolute value squared of the Wilson coefficients of the operators,
which can be obtained by matching the full theory onto SCET.
It is obtained by subtracting the IR divergences in the $\overline{MS}$ scheme
from the UV renormalized amplitudes of the full theory.
At the LO, the hard function $H$ is normalized to $1$.
In general, it is related to the amplitudes of the full theory, using
\begin{eqnarray}
 \lambda_{0,ij} H^{(0)}_{IJ} &=& \frac{1}{\la c_I|c_I\ra \la
c_J
| c_J \ra} \la c_I | \mathcal{M}^{(0)}_{\rm ren} \ra \la
\mathcal{M}^{(0)}_{\rm ren} | c_J \ra,
\nn
\\
\lambda_{0,ij} H^{(1)}_{IJ} &=& \frac{1}{\la c_I|c_I\ra \la
c_J |
c_J \ra} \left(\la c_I | \mathcal{M}^{(1)}_{\rm ren} \ra
\la
\mathcal{M}^{(0)}_{\rm ren} | c_J \ra+\la c_I |
\mathcal{M}^{(0)}_{\rm ren} \ra \la
\mathcal{M}^{(1)}_{\rm ren} | c_J \ra\right),
\label{eqs:hardmatrix}
\end{eqnarray}
where $|\mathcal{M}_{\rm ren} \ra$ are obtained by
subtracting the IR divergences in the $\overline{\rm MS}$ scheme
from the UV  renormalized amplitudes of the full theory~\cite{Manohar:2003vb,Idilbi:2005ky,Ahrens:2010zv}.
At NLO, in practice, it is necessary to calculate the one-loop on-shell Feynman diagrams of this process, as shown in Fig. \ref{fig:virtual}.
Using the one-loop results in Refs. \cite{Schmidt:1997wr,Ravindran:2002dc},
we get the hard functions at NLO  as follows
\begin{eqnarray}
\label{hard_function}
H_{gg}(\mu_{h})&=&1+\frac{\alpha_{s}(\mu_{h})}{4\pi}
\biggl\{-3 N_c\mathrm{ln}^2\left(\frac{\mu_{h}^2}{M_H^2}\right) \nn\\&&
+\left[\gamma^{H,0}_{gg}-2 N_c \left(\ln \left(\frac{M_H^2}{\hat{s}}\right)+\ln \left(\frac{M_H^2}{-\hat{t}}\right)+\ln \left(\frac{M_H^2}{-\hat{u}}\right)\right)\right]~\mathrm{ln}\left(\frac{\mu _{h}^2}{M_H^2}\right) +c^{H,gg}_1 \biggr\},\\
\label{hardf_function}
H_{gq}(\mu_{h})&=&1+\frac{\alpha_{s}(\mu_{h})}{4\pi}
\biggl[\left(\frac{1}{N_c}-2 N_c\right) \mathrm{ln}^2\left(\frac{\mu _{h}^2}{M_H^2}\right) \nn\\&&
+\frac{N_c^2 \left(-6 \ln \left(\frac{M_H^2}{\hat{s}}\right)-6 \ln \left(\frac{M_H^2}{-\hat{u}}\right)+13\right)+6 \ln \left(\frac{M_H^2}{-\hat{t}}\right)-4 N_c n_f+9}{3 N_c}
   ~\mathrm{ln}\left(\frac{\mu_{h}^2}{M_H^2}\right)+c^{H,gq}_1 \biggr],\nn\\
\end{eqnarray}
with
\begin{eqnarray}
 c^{H,gg}_1 &=& 3 \left[4 \text{Li}_2\left(1-\frac{M_H^2}{\hat{s}}\right)+4 \text{Li}_2\left(\frac{\hat{u}}{M_H^2}\right)+4 \text{Li}_2\left(\frac{\hat{t}}{M_H^2}\right)
  +\ln^2\left(\frac{M_H^2}{\hat{s}}\right)-\ln ^2\left(\frac{M_H^2}{-\hat{u}}\right) \right.\nn\\&&\left.
   -\ln ^2\left(\frac{M_H^2}{-\hat{t}}\right)-2 \ln \left(\frac{\hat{s}}{M_H^2}\right) \ln
   \left(\frac{-\hat{u}}{M_H^2}\right)-2 \ln \left(\frac{\hat{s}}{M_H^2}\right) \ln \left(\frac{-\hat{t}}{M_H^2}\right)
   -2 \ln \left(\frac{-\hat{u}}{M_H^2}\right)\ln   \left(\frac{-\hat{t}}{M_H^2}\right) \right.\nn\\&&\left.
    +4 \ln \left(\frac{-\hat{u}}{M_H^2}\right) \ln \left(1-\frac{\hat{u}}{M_H^2}\right)
   +4 \ln \left(\frac{-\hat{t}}{M_H^2}\right)\ln   \left(1-\frac{\hat{t}}{M_H^2}\right)  +\frac{25\pi ^2}{6} \right]  \nn\\&&
   +\frac{2}{3}\frac{(N_c-n_f)M_H^2 [M_H^2(\hat{s} \hat{t}+\hat{s} \hat{u}+\hat{t} \hat{u} ) + \hat{s} \hat{t} \hat{u}]}{\left(M_H^8+\hat{s}^4+\hat{t}^4+\hat{u}^4\right)} + 22
\end{eqnarray}

\begin{eqnarray}
c^{H,gq}_1&=&N_c V_1 + \frac{1}{N_c} V_2 + n_f V_3+V_4,
\end{eqnarray}
where
\begin{eqnarray}
 V_1 &=& 4 \text{Li}_2\left(1-\frac{\hat{t}}{M_H^2}\right)
 +2 \text{Li}_2\left(1-\frac{\hat{u}}{M_H^2}\right)+2\text{Li}_2\left(1-\frac{\hat{s}}{M_H^2}\right)
 -\frac{13}{3}\ln \left(\frac{-\hat{t}}{M_H^2}\right)\nn\\&&
 -\ln^2\left(\frac{M_H^2}{-\hat{u}}\right)
 -2 \ln \left(\frac{-\hat{t}}{M_H^2}\right) \ln \left(\frac{-\hat{u}}{M_H^2}\right)
 -2 \ln \left(\frac{\hat{s}}{M_H^2}\right) \ln \left(\frac{-\hat{t}}{M_H^2}\right)   \nn\\&&
+4  \ln \left(1-\frac{\hat{t}}{M_H^2}\right) \ln \left(\frac{-\hat{t}}{M_H^2}\right)
+2\ln \left(1-\frac{\hat{u}}{M_H^2}\right) \ln \left(\frac{-\hat{u}}{M_H^2}\right)+\frac{80}{9},\\
V_2 &=& -2\text{Li}_2\left(1-\frac{\hat{s}}{M_H^2}\right)-2 \text{Li}_2\left(1-\frac{\hat{u}}{M_H^2}\right)
 + \ln^2\left(\frac{M_H^2}{-\hat{t}}\right)-\ln^2\left(\frac{M_H^2}{\hat{s}}\right)-3 \ln \left(\frac{-\hat{t}}{M_H^2}\right)\nn\\&&
 + 2 \ln \left(\frac{\hat{s}}{M_H^2}\right) \ln \left(\frac{-\hat{u}}{M_H^2}\right)
-2\ln \left(1-\frac{\hat{u}}{M_H^2}\right) \ln \left(\frac{-\hat{u}}{M_H^2}\right) - \frac{\pi^2}{6}+8,\\
V_3 &=& \frac{4}{3} \ln \left(\frac{-\hat{t}}{M_H^2}\right)-\frac{20}{9}, \\
V_4 &=& \frac{10}{3}\frac{-\hat{t}(\hat{u}+\hat{s})}{\hat{u}^2+\hat{s}^2}+22.
\end{eqnarray}

Our results of hard functions are consistent with the results in Ref \cite{Jouttenus:2013hs}.
The hard functions at the other scales can be obtained by evolution of renormalization group (RG) equations.
The RG equations for hard functions are governed by the anomalous-dimension matrix,  which
has been calculated in Refs. \cite{Becher:2009kw,Ferroglia:2009ep,Ferroglia:2009ii,Becher:2009cu,Becher:2009qa,Ahrens:2012qz}.
In our case,  the RG equations for hard functions are given by
\begin{eqnarray}\label{eqs:hardupRG}
\frac{d}{d~\mathrm{ln}\mu_{h}}H_{gg}(\mu_{h})&=&\biggl[
3 \gamma_{\rm cusp}\left(\mathrm{ln}\frac{\hat{s}}{\mu_{h}^2}+\mathrm{ln}\frac{-\hat{t}}{\mu_{h}^2}+\mathrm{ln}\frac{-\hat{u}}{\mu_{h}^2}\right)
+2\gamma_{gg}^H \biggr ]H_{gg}(\mu_{h}),\\
\label{eqs:harddnRG}
\frac{d}{d~\mathrm{ln}\mu_{h}}H_{gq}(\mu_{h})&=&
\biggl[3\gamma_{\rm cusp}\left( \mathrm{ln}\frac{\hat{s}}{\mu_{h}^2}+\mathrm{ln}\frac{-\hat{u}}{\mu_{h}^2}-\frac{1}{9}\mathrm{ln}\frac{-\hat{t}}{\mu_{h}^2} \right)
+2\gamma_{gq}^H \biggr ]H_{gq}(\mu_{h}),
\end{eqnarray}
with
\begin{eqnarray}
  2\gamma_{gg}^H &=& 2 \gamma_{gg}^V -3\frac{\beta(\alpha_s)}{\alpha_s},\quad \gamma_{gg}^{H,0}=0, \\
  2\gamma_{gq}^H &=& 2 \gamma_{gq}^V  -3\frac{\beta(\alpha_s)}{\alpha_s},\quad \gamma_{gq}^{H,0}=-6C_F+2\beta_0,
\end{eqnarray}
where $\gamma_{\rm cusp}$ is  the universal anomalous-dimension function related to the cusp anomalous dimension
of Wilson loops with lightlike segments~\cite{Korchemsky:1987wg,Korchemsky:1988hd,Korchemskaya1992169},
while $\gamma_{gg}^V$ and $\gamma_{gq}^V$ control the single-logarithmic evolution.
Their explicit expressions are shown in Ref.~\cite{Becher:2009qa}. In the following,  all anomalous dimensions are expanded in
unit of $\alpha_s/4\pi$, for example, $\gamma_{\rm cusp}(\alpha)=\frac{\alpha_s}{4 \pi}\Gamma_0+(\frac{\alpha_s}{4 \pi})^2\Gamma_1+\mathcal{O}(\alpha_s^3)$.

Solving the RG equations, the hard function at an arbitrary scale $\mu$ are given by:
\begin{eqnarray}\label{hardgg}
H_{gg}(\mu)&=& \left( \frac{\alpha_s(\mu_h)}{\alpha_s(\mu)} \right)^3
\mathrm{exp}\left[ 18S(\mu_{h},\mu)-2a_{gg}^V(\mu_{h},\mu)\right]\bigg( \frac{\hat{s}\hat{t}\hat{u}
}{\mu_{h}^6} \bigg)^{-3a_{\Gamma}(\mu_{h},\mu)} H_{gg}(\mu_{h}),\\
H_{gq}(\mu)&=& \left( \frac{\alpha_s(\mu_h)}{\alpha_s(\mu)} \right)^3
\mathrm{exp}\left[ \frac{34}{3}S(\mu_{h},\mu)-2a_{gq}^V(\mu_{h},\mu)\right]\bigg( \frac{(\hat{s})^9(-\hat{u})^9/(-\hat{t})}{\mu_{h}^{34}} \bigg)^{-\frac{1}{3}a_{\Gamma}(\mu_{h},\mu)} H_{gq}(\mu_{h}),\nn\\
\end{eqnarray}
where $S(\mu_{h},\mu)$ and $a_{gi}^V$ are defined as~\cite{Becher:2006mr}
\begin{eqnarray}
S(\mu_{h},\mu)&=&-\int_{\alpha_s(\mu_{h})}^{\alpha_s(\mu)}d\alpha \frac{\gamma_{\rm cusp}(\alpha)}{\beta(\alpha)}
 \int_{\alpha_s(\mu_{h})}^{\alpha}\frac{d\alpha^{\prime}}{\beta(\alpha^{\prime})},\\
a_{gi}^V(\mu_{h},\mu)&=&-\int_{\alpha_s(\mu_{h})}^{\alpha_s(\mu)}d\alpha
\frac{\gamma_{gi}^V(\alpha)}{\beta(\alpha)}.
\end{eqnarray}
The general hard function up to $\mathcal{O}(\alpha_s^2)$ can be written as
\begin{align}
\tilde{H}_{ij}  &= 1
+ \left( \frac{\alpha_s}{4 \pi} \right) \left\{ -\rho_{ij}\Gamma_0 \frac{L^2_{H}}{2} - \tilde{\gamma}_0^{H_{ij}} L_{H} + c_1^H \right\}
\nonumber\\
& + \left( \frac{\alpha_s}{4 \pi} \right)^2
\bigg\{ \left(\rho_{ij} \Gamma_0  \right)^2 \frac{L^4_{H}}{8}
+  \left( \beta_0+   3 \tilde{\gamma}_0^{H_{ij}} \right) \rho_{ij}\Gamma_0 \frac{L^3_{H}}{6}
\nonumber\\
& \qquad + \left[ \tilde{\gamma}_0^{H_{ij}}(\beta_0+\tilde{\gamma}_0^{H_{ij}}) - \rho_{ij} \Gamma_1 - \rho_{ij} \Gamma_0 c_1^H \right] \frac{L^2_{H}}{2}
\nonumber \\
& \qquad +
\left[ - c_1^H (\beta_0 + \tilde{\gamma}_0^{H_{ij}}) - \tilde{\gamma}_1^{H_{ij}} \right] L_H + c_2^H \bigg\}.
\end{align}
The coefficients in the above equation can be obtained from Refs.\cite{Gehrmann:2011aa,Becher:2013vva}.
Here, $L_H=\mathrm{ln}\left(\frac{\mu_{h}^2}{M_H^2}\right),
\rho_{gg}=9/2, \rho_{gq}=17/6$, $\tilde{\gamma}^{{H_{gg}}}(\alpha_s) = 3\gamma^g(\alpha_s)+\frac{C_A}{2}\Gamma \ln\frac{\hat{s}\hat{t}\hat{u}}{\mu_h^6} - \frac{3\beta(\alpha_s)}{2\alpha_s}$, $\tilde{\gamma}^{{H_{gq}}}(\alpha_s) = 2\gamma^g(\alpha_s)+\gamma^q+\frac{C_A}{18}\Gamma \ln\frac{\hat{s}^9(-\hat{u})^9}{-\hat{t}\mu_h^{34}} - \frac{3\beta(\alpha_s)}{2\alpha_s}$.

\subsection{Jet function}
The jet function of gluon $J_g(p^2)$  is defined as
\begin{equation}
 \langle  0 |\, {{\cal A}_{J}^a}_\perp^\mu( x) {{\cal A}_{J}^b}_\perp^\nu(0)\,   | 0 \rangle =
(-g_\perp^{\mu\nu})\, \delta^{ab} \, g_s^2\, \int \frac{\mathrm{d}^4
p}{(2\pi)^3}\,\theta(p^0)\, J_g(p^2) \,e^{-i p x}\, .
\end{equation}
These collinear gluon operators  have nonvanishing matrix elements only for intermediate collinear states. Thus, this jet function
can be considered  as the result of integrating out the collinear modes at the scale $\mu_j$. Equivalently, we can extract the jet function from the imaginary part of the time-ordered product of collinear fields
\begin{equation}
\frac{1}{\pi} {\rm Im} \left[ i \int \mathrm{d}^4x\, e^{i p x}  \langle 0 | \,  {\bm T}\left\{ {{\cal A}_{J}^a}_\perp^\mu( x) {{\cal A}_{J}^b}_\perp^\nu(0) \right \} \, | 0 \rangle \right] =(-g_\perp^{\mu\nu})\, \delta^{ab}\, g_s^2\, {J}_g(p^2)\,.
\end{equation}
The RG evolution of the jet function is given
by
\begin{equation}
 \frac{dJ_i(p^2,\mu)}{d\ln\mu} = \left( -2 \gamma_{\rm cusp}
\ln\frac{p^2}{\mu^2} - 2 \gamma^J_i \right)J_i(p^2,\mu)
+2\gamma_{\rm cusp}\int^{p^2}_0
dq^2\,\frac{J_i(p^2,\mu)-J_i(q^2,\mu)}{p^2-q^2}
\end{equation}
with $i=g,q$.
To solve this integro-differential evolution equation, we use the Laplace
transformed jet function~\cite{Becher:2006mr}
\begin{equation}\label{eqs:jetfunction}
 \widetilde{j_i}(\ln\frac{Q^2}{\mu^2},\mu)=\int^\infty_0
dp^2\,\exp(-\frac{p^2}{Q^2e^{\gamma_E}}) J_i(p^2,\mu),
\end{equation}
which satisfies the the RG equation
\begin{equation}\label{eqs:jetRG}
 \frac{d}{d\ln\mu}\widetilde{j_i}(\ln\frac{Q^2}{\mu^2},\mu)=\left(-2C_i\gamma_{\rm cusp}
\ln\frac{Q^2}{\mu^2}-2\gamma^{J_i}\right)\widetilde{j_i}(\ln\frac{Q^2}{\mu^2},\mu).
\end{equation}
The Laplace transformed jet
function $\widetilde{j_i}(L,\mu)$ at NNLO \cite{Becher:2006qw,Becher:2010pd} is
\begin{eqnarray}\label{jet function_nlo}
\widetilde{j_i}(L,\mu)&=&1+\frac{\alpha_s}{4\pi}\bigg( \frac{C_i\Gamma_0}{2}L^2+\gamma^{J_i}_0L+c^{J_i}_1\bigg) \nn\\
&&+\biggl(\frac{\alpha_s}{4\pi}\biggr)^2\bigg\{\frac{C_i^2\Gamma_0^2}{8}L^4
+(\frac{\gamma^{J_i}_0C_i\Gamma_0}{2}-\frac{\beta_0C_i\Gamma_0}{6})L^3
+\half[C_i\Gamma_1+(\gamma^{J_i}_0-\beta^0)\gamma^{J_i}_0+c^{J_i}_1C_i\Gamma_0]L^2\nn\\
&&+[\gamma^{J_i}_1+(\gamma^{J_i}_0-\beta_0)c^{J_i}_1]L
+c^{J_i}_2\biggr\}
\end{eqnarray}
with

\begin{eqnarray}
c_1^{J_q} &=&\left( 7- \frac{2}{3}\pi^2 \right)C_F, \nn \\
c_1^{J_g} &=&\left( \frac{67}{9} -\frac{2}{3}\pi^2\right)C_A-\frac{20}{9} n_f T_F, \nn \\
  c_2^{J_q} &=& \left(\frac{205}{8}-\frac{97\pi^2}{12}+\frac{61\pi^4}{90}-6\zeta_3\right)C_F^2
  +\left(\frac{53129}{648}-\frac{155\pi^2}{36}-\frac{37\pi^4}{180}-18\zeta_3\right)C_F C_A \nn\\
  &+&\left(-\frac{4057}{162}+\frac{13\pi^2}{9} \right)C_F n_f T_F, \\
  c^{J_g}_2 &=& C_A^2 \left(\frac{20215}{162}-\frac{362 \pi ^2}{27}-\frac{88 \zeta_3}{3}+\frac{17 \pi
   ^4}{36}\right)+  C_A n_f T_F \left(-\frac{1520}{27}+\frac{134 \pi^2}{27}-\frac{16 \zeta_3}{3}\right) \nn\\
   &+&   C_F n_f T_F\left(-\frac{55}{3}+16 \zeta_3\right) +  n_f^2 T_F^2\left(\frac{400}{81}-\frac{8 \pi ^2}{27}\right) \,.
\end{eqnarray}
Following the approach shown in Ref \cite{Becher:2006nr},   the RG-improved  jet function at an arbitrary scale $\mu$ can be
obtained
\begin{equation}
 {J_i}(p^2,\mu)=\exp \bigl[ -4C_iS(\mu_j,\mu)+2a^{J_i}(\mu_j,\mu)
\bigr] \widetilde{j}(\partial_{\eta_j}, \mu_j )  \frac{1}{p^2} \left(
\frac{p^2}{\mu^2_j}\right)^{\eta_j}
\frac{e^{-\gamma_E
\eta_j}}{\Gamma(\eta_j)},
\label{jet function}
\end{equation}
where $\eta_j=2 a_\Gamma(\muj,\mu),C_g=3,C_q=4/3$.
\subsection{Soft function}
The soft function $\mathcal{S}(k,\mu)$, which describes soft interactions between all colored particles, can be calculated perturbatively in SCET.
For the gg channel, the soft function is defined as
\begin{equation}\label{sfdefine}
 \langle 0 |\bar{\textbf{T}}[Y_J^{\dagger}Y_2^{\dagger}Y_1^{\dagger}(x_-)] \textbf{T} [Y_1 Y_2 Y_J(0)]|0 \rangle = \int_0^{\infty} \mathrm{d} k_+ \, e^{-i k_+ ( { \bar{n}_J} \cdot x)/2} S_{gg}(k_+)
\end{equation}
For our threshold resummation at large $p_T$, the soft function becomes scaleless in dimensional regularization,
which is consistent with the regularization scheme of hard function and jet function.
Actually, we only need to
calculate the emission diagrams, using dimensional regularization.
The soft function
$\mathcal{S}(k,\mu)$, similar to the jet function,
satisfies the RG equation \cite{Becher:2006nr,Becher:2007ty}
\begin{eqnarray}
\frac{d }{d \ln \mu}\mathcal{S}(k,\mu)=\biggl[ -4C_{gi}\gamma_{\rm cusp}\ln\frac{k}{\tilde{\mu}}+2\gamma^S \biggr]\mathcal{S}(k,\mu)
+4C_{gi}\gamma_{\rm cusp}\int_0^k d k^{\prime}\frac{\mathcal{S}(k,\mu)-\mathcal{S}(k^{\prime},\mu)}{k-k^{\prime}}.
\end{eqnarray}
According to the method shown in Refs.~\cite{Becher:2006nr,Becher:2007ty},  the RG-improved  soft function can be given as
\begin{eqnarray}\label{eqs:soft_function}
\mathcal{S}(k,\mu)=\exp \bigl[ -4C_{gi}S(\mu_s,\mu)-2a^S(\mu_s,\mu)
\bigr]\widetilde{s} (\partial_{\eta_s},\mu_s)\frac{1}{k}\biggl(
\frac{k}{\tilde{\mu}_s} \biggr)^{\eta_s}
\frac{e^{-\gamma_E\eta_s}}{\Gamma(\eta_s)},
\end{eqnarray}
where $\eta_s=2a_{\Gamma}(\mu_s,\mu),C_{gg}=3/2,C_{gq}=3/2$,  and the Laplace transformed soft
function $\widetilde{s}(L,\mu)$ at NNLO is given by \cite{Becher:2012za}
\begin{eqnarray}\label{eqs:soft_function_nlo}
\widetilde{s}(L,\mu)&=&1+\frac{\alpha_s}{4\pi}\bigg( 2C_{gi}\Gamma_0 L^2-2\gamma^S_0 L+ c^S_1 \bigg) \nn\\
&&+ \biggl(\frac{\alpha_s}{4\pi}\biggr)^2
 \biggl\{  2C_{gi}^2 \Gamma_0^2L^4
 + ( -4\gamma^S_0C_{gi}\Gamma_0-\frac{4\beta_0C_{gi}\Gamma_0}{3})L^3  \nn\\
&& + \left[ 2C_{gi}\Gamma_1+2(\gamma^{S}_{0}+\beta_0)\gamma^S_0+2c^S_1C_{gi}\Gamma_0 \right]L^2
-2\left[ \gamma^S_1+(\gamma^S_0+\beta_0) c^S_1 \right]L
 + c^S_2 \biggr\} .
\end{eqnarray}
with
\begin{equation}
c_1^S=\pi^2 C_A/2
\end{equation}
and
\begin{equation}
    c_2^S=\left(\frac{1214}{81}+\frac{335\pi^2}{108}-\frac{11\zeta_3}{9}-\frac{41\pi^4}{120}\right)C_A^2
    +\left(-\frac{328}{81}-\frac{25\pi^2}{27}+\frac{4\zeta_3}{9}\right)C_A n_f T_F.
\end{equation}

\subsection{Scale independence}
In the factorization formalism, we have introduced the hard function, jet function and soft function.
Each of them is evaluated at a scale to make the perturbatvie expansion reliable, and then evolved
to a common scale $\mu_F$ in the PDFs. Therefore, it is important to check the scale independence of the final results.
In fact, after expanding the exponent in Eq. (\ref{hardgg}), we can find the dependence on the
intermediate scale $\mu_{h}$ cancel each other up to $\mathcal{O}(\alpha_s)$.
For the  dependence of the jet scale, it is more complicate due to the appearance of the partial derivative operator
and the delta function in the expansion of the jet function
\begin{equation}\label{eqs:expan}
    \frac{1}{p^2} \left(\frac{p^2}{\mu^2_j}\right)^{\eta_j}=\frac{\delta (p^2)}{\eta_j}
    +\left[\frac{1}{p^2}\right]_{\star}^{[p^2,\mu_j^2]}
    +\eta_j\left[\frac{\ln (p^2/\mu_j^2)}{p^2}\right]_{\star}^{[p^2,\mu_j^2]}+\mathcal{O}(\eta_j^2).
\end{equation}
The star distribution is defined as
\begin{equation}\label{star}
   \int_0^{Q^2}\!dp^2\,\left[ \frac{1}{p^2}
   \left( \frac{p^2}{\mu^2} \right)^\eta \right]_{\star}\,f(p^2)
   = \int_0^{Q^2}\!dp^2\,\frac{f(p^2)-f(0)}{p^2}
    \left( \frac{p^2}{\mu^2} \right)^\eta
    + \frac{f(0)}{\eta} \left( \frac{Q^2}{\mu^2} \right)^\eta ,
\end{equation}
where $f(p^2)$ is a smooth test function, and the $f(0)$ subtraction term
is needed only if $\eta<0$.
The scale independence happens for the jet function only in the sense of the integration over $p^2$.
For the dependence of the  soft scale, it is the same as the jet function, and we do not discuss it here.

Now we begin  to  discuss the dependence of the  the final results.
Using the hadronic threshold definition in Eq. (\ref{eqs:s4}) and the cross section near the threshold in Eq. (\ref{eqs:facmain}),
we have
\begin{eqnarray}\label{eqs:scaleind}
\frac{d\sigma}{dS_4dy} \propto && \int dx_a dx_b\int dp_1^2 \int dk^+\frac{1}{\hat{s}}
f_{i/P_a}(x_a,\mu) f_{j/P_b}(x_b,\mu) H_{ij}(\mu)  \nn\\&&
J(p_1^2,\mu) S(k^+,\mu) \delta(S_4-(-\hat{t})(1-x_a)-(-\hat{u})(1-x_b)-p_1^2-2k^+E_1),
\end{eqnarray}
where we have changed the integration variables $d\hat{t}d\hat{u}$  to $dS_4 dy$.
From this equation, we can see clearly the connection between the threshold region of the whole system, represented by $S_4$,
and those of the parts of the system, represented by $(1-x_a),(1-x_b), p_1^2, k^+$ respectively.
For simplifying the convolution form,
using the Laplace transformation, the above equation can be written as
\begin{equation}
    \frac{d\tilde{\sigma}}{dQ^2dy}=\int_0^{\infty} dS_4 \exp \left(-\frac{S_4}{Q^2e^{\gamma_E}}\right)\frac{d\sigma}{dS_4dy}.
\end{equation}
The Laplace transformed PDFs near the end point are given by
\begin{equation}
\tilde{f}_{i/P}(\tau,\mu)=\int_0^1 dx \exp \left(-\frac{1-x}{\tau e^{\gamma_E}}\right) f_{i/P_a}(x,\mu),
\end{equation}
which satisfies RG equation
\begin{equation}
 \frac{d}{d\ln\mu}\tilde{f}_{i/P}(\tau,\mu)=\left(2C_i\gamma_{\rm cusp}
\ln (\tau) + 2\gamma^{\phi i}\right)\tilde{f}_{i/P}(\tau,\mu).
\end{equation}
The variable $\tau$ in the Laplace transformed PDF is given by
\begin{equation}
    \tau_a = \frac{Q^2}{-\hat{t}}{\rm~~~~for~~~~} \tilde{f}_{i/P_a}(\tau_a,\mu),\quad {\rm and} \quad
    \tau_b = \frac{Q^2}{-\hat{u}}{\rm~~~~for~~~~} \tilde{f}_{j/P_b}(\tau_b,\mu).
\end{equation}
Using the relations between the  anomalous dimensions presented in Ref. \cite{Becher:2009qa},
we  have
\begin{equation}
    \frac{d}{d\ln \mu}\left[\tilde{f}_{g/P_a}(\tau_a,\mu) \tilde{f}_{g/P_b}(\tau_b,\mu) \hat{\sigma}_{gg,B}(\mu) H_{gg} (\mu)
    \widetilde{j_g}(\ln\frac{Q^2}{\mu^2},\mu)\widetilde{s}(\ln \frac{Q^2\sqrt{\hat{s}}}{\mu\sqrt{\hat{u}\hat{t}}},\mu)\right]=0
\end{equation}
and
\begin{equation}
    \frac{d}{d\ln \mu}\left[\tilde{f}_{g/P_a}(\tau_a,\mu) \tilde{f}_{q/P_b}(\tau_b,\mu) \hat{\sigma}_{gq,B}(\mu) H_{gq} (\mu)
    \widetilde{j_q}(\ln\frac{Q^2}{\mu^2},\mu)\widetilde{s}(\ln \frac{Q^2\sqrt{\hat{s}}}{\mu\sqrt{\hat{u}\hat{t}}},\mu)\right]=0,
\end{equation}
which show the scale independence of the cross section.

\subsection{Final RG-improved differential cross section}
Combining the RG-improved hard, soft and jet functions,
and using the identities \cite{Becher:2008cf}
\begin{align}
 a_\Gamma(\mu_1,\mu_2) +a_\Gamma(\mu_2,\mu_3) &= a_\Gamma(\mu_1,\mu_3)\,, \nonumber \\
 S (\mu_1, \mu_2) + S(\mu_2, \mu_3) &= S (\mu_1, \mu_3) + \ln \frac{\mu_1}{\mu_2} a_{\Gamma} (\mu_2, \mu_3) \,, \nonumber \\
  f (\partial_{\eta}) X^{\eta}  &= X^{\eta} f (\ln X +
  \partial_{\eta})\,,
\end{align}
we get the resummed differential cross
section for the Higgs boson and a jet associated  production
\begin{eqnarray}\label{finalresum}
\frac{d\hat{\sigma_{ij}}^{\rm thres}}{d\hat{t}d\hat{u}} &=&
\sum_{ij}\frac{\lambda_{0,ij}(\mu_h)}{16\pi\hat{s}^2 } \nn\\
&&
\mathrm{exp}\left[4 \rho_{ij} S(\mu_{h},\mu)-2a_{ij}^V(\mu_{h},\mu)\right]\tilde{H}_{ij}(\mu_{h}) \nn\\
&&\exp \bigl[ -4C_i S(\mu_j,\mu)+2a^{J_i}(\mu_j,\mu)\bigr]\left( \frac{M_H^2}{\mu^2_j} \right)^{\eta_j} \nn\\
&& \exp \bigl[ -4C_{gi}S(\mu_s,\mu)-2a^S(\mu_s,\mu)\bigr]
\biggl( \frac{M_H^2\sqrt{\hat{s}}}{\mu_s \sqrt{\hat{t}\hat{u}}} \biggr)^{\eta_s}  \nn\\
&& \widetilde{j}(\partial_{\eta}+L_j , \mu_j )\widetilde{s}(\partial_{\eta}+L_s,\mu_s)
 \frac{1}{s_4}\left( \frac{s_4}{M_H^2} \right)^{\eta}
 \frac{e^{-\gamma_E\eta}}{\Gamma(\eta)},
\end{eqnarray}
where $C_g=3,C_q=4/3,C_{gg}=3/2,C_{gq}=3/2,
\eta=\eta_j+\eta_s$, $L_j=\ln (M_H^2/\mu^2_j)$ and $L_s=\ln (M_H^2\sqrt{\hat{s}})/(\mu_s \sqrt{\hat{t}\hat{u}})$.

In order to compare with the fixed-order results,  setting $\mu_{h}=\mu_{j}=\mu_{s}=\mu$,
we expand the above results  up to $\mathcal{O}(\alpha_s^2)$
\begin{eqnarray}\label{eqs:singular}
\left( \frac{\lambda_{0,ij}}{16\pi\hat{s}^2 } \right)^{-1}\frac{d\hat{\sigma}_{ij}^{\rm thres}}{d\hat{t}d\hat{u}} &=&\delta(s_4)
+\frac{\alpha_s}{4\pi}\biggl\{A_2 D_2+A_1 D_1 + A_0 \delta(s_4) \biggr\} \nn\\
&&\hspace{-3.0cm}+\bigg(\frac{\alpha_s}{4\pi}\bigg)^2 \bigg\{ B_4 D_4+ B_3 D_3+ B_2 D_2+ B_1 D_1 + B_0 \delta(s_4)\bigg\},
\end{eqnarray}
with
\begin{equation}\label{eqs:D}
    D_n=\biggr[\frac{\ln^{n-1} (s_4/M_H^2)}{s_4} \biggl]_{+},
\end{equation}
where the coefficients of  $A_n$ and $B_n$  are given by
\begin{eqnarray}
  A_2 &=& (C_i+4C_{gi})\Gamma_0, \\
  A_1 &=& (C_iL_j+4C_{gi}L_s)\Gamma_0 + \gamma^{J_i}_0-2\gamma^S_0,  \\
  A_0 &=& \left[\frac{1}{2} C_i L_j^2 + 2C_{gi}L_s^2-\frac{\pi^2}{12}(C_i+4C_{gi})-\frac{\rho_{ij}}{2}L_{H}^2 \right]\Gamma_0  +
  \gamma^{J_i}_0 L_j - 2\gamma^S_0 L_s\nn\\
   &+& \tilde{\gamma}_0^{H_{ij}} L_{H}  +c^{J_i}_1+ c^S_1+ c^H_1, \\
  B_4 &=& \frac{A_2^2}{2}, \\
  B_3 &=& \frac{3}{2}A_2A_1-\frac{1}{2}\beta_0(C_i+8C_{gi})\Gamma_0, \\
  B_2 &=& A_1^2+A_2\left(A_0-\frac{\pi^2}{6}A_2\right)-\beta_0[(C_iL_j+8C_{gi}L_s)\Gamma_0+\gamma^{J_i}_0-4\gamma^S_0]+(C_i+4C_{gi})\Gamma_1,\\
  B_1 &=& \zeta_3A_2^2 + A_1\left(A_0-\frac{\pi^2}{6}A_2\right)    \nn\\
  &-&\beta_0\left[\frac{\Gamma_0}{2}(C_i L_j^2+8C_{gi}L_s^2)-\frac{\pi^2 \Gamma_0}{12}(C_i+8C_{gi})
  +\gamma^{J_i}_0L_j-4\gamma^S_0L_s+c^{J_i}_1+2c^S_1\right] \nn\\
  &+&(C_i L_j + 4 C_{gi} L_s )\Gamma_1 + \gamma^{J_i}_1 - 2 \gamma^S_1, \\
\label{b0}  B_0 &=& \frac{A_0^2}{2}-\frac{\pi^4}{720}A_2^2-\frac{\Gamma_0^2}{12}(C_i L_j + 4C_{gi} L_s)[\pi^2(C_i L_j + 4C_{gi}L_s)-12\zeta_3(C_i+4C_{gi})] \nn  \\
  &-&\frac{\Gamma_0}{6}(\gamma^{J_i}_0-2\gamma^S_0)[\pi^2(C_i L_j + 4C_{gi} L_s)-6\zeta_3(C_i+4C_{gi})]+\rho_{ij} \tilde{\gamma}_0^{H_{ij}} \Gamma_0 L_H^3 \nn\\
  &-&\frac{\pi^2}{12}[\Gamma_1(C_i+4C_{gi})+(\gamma^{J_i}_0-2\gamma^S_0)^2]+\frac{\Gamma_1}{2}(C_i L_j^2 + 4C_{gi}L_s^2-\rho_{ij}  L_H^2) \nn\\
  &+&(\gamma^{J_i}_1L_j-2\gamma^S_1 L_s-L_H \tilde{\gamma}_1^{H_{ij}}-2c_1^H \tilde{\gamma}_0^{H_{ij}} L_H )-\frac{1}{2}({c^{J_i}_1}^2+{c^S_1}^2+{c^H_1}^2)+c^{J_i}_2+c^S_2+c^H_2 \nn\\
  &+&\frac{\beta_0}{12}\big\{\Gamma_0[C_i(-2L_j^3+\pi^2L_j-4\zeta_3)+8C_{gi}(-2L_s^3+\pi^2L_s-4\zeta_3) -2\rho_{ij} L_H^3
  ] \nn\\
  &+&\pi^2(\gamma^{J_i}_0-4\gamma^S_0)-6(\gamma^{J_i}_0 L_j^2-4\gamma^S_0 L_s^2+2c^J_1L_j+4c^S_1 L_s)-12 c_1^H L_H+6 \tilde{\gamma}_0^{H_{ij}} L_H^2 \big\},
\end{eqnarray}
with $\zeta_3=1.20206\cdots$.
We find that the coefficient $A_{2,1,0}$ agree with the NLO results in Ref.~\cite{Ravindran:2002dc}.

In order to obtain the best possible precise predictions, we combine our resummed result with the nonsingular terms up to NLO in fixed-order
perturbative calculations, and  the RG-improved differential cross section are given by
\begin{equation}\label{eqs:match}
\frac{d\hat{\sigma}_{\rm Resum}}{d\hat{t}d\hat{u}}=\frac{d\hat{\sigma}^{\rm thres}}{d\hat{t}d\hat{u}}
  +\left( \frac{d\hat{\sigma}_{\rm NLO}}{d\hat{t}d\hat{u}} -\frac{d\hat{\sigma}^{\rm thres}}{d\hat{t}d\hat{u}}  \right)|_{\rm expanded~to~NLO} ,
\end{equation}
where the NLO results can be obtained by the modified Monte Carlo programs MCFM \cite{Campbell:2010ff} or HNNLO  \cite{Catani:2007vq,Grazzini:2008tf,Grazzini:2013mca}.
Near the threshold regions, the expansion of the resummed results approaches the fixed-order one so that the terms in the bracket almost vanishes and the threshold contribution  dominates. In the regions far from the threshold limit, the fixed-order contribution dominates and the resummation effects are not important.

\section{Numerical Discussion}
\label{sec:nume}
In this section, we discuss the relevant numerical results.
The Higgs boson mass and top quark mass are chosen as $125.6$ and
$173.2$ ~GeV \cite{CDF:2013jga}, respectively.
The CTEQ6M PDF sets are used
throughout our numerical calculations.
And the factorization scale is set at $M_H$ unless special statement.
There are three new scales, i.e., $\mu_{h},\mu_j,\mu_s$, introduced in the SCET formalism.
They should be properly chosen so that the corresponding hard functions, jet function and soft function have stable numerical results,
which means each function should  not contain large logarithms at the chosen scale.
\begin{figure}
  \includegraphics[width=0.49\linewidth]{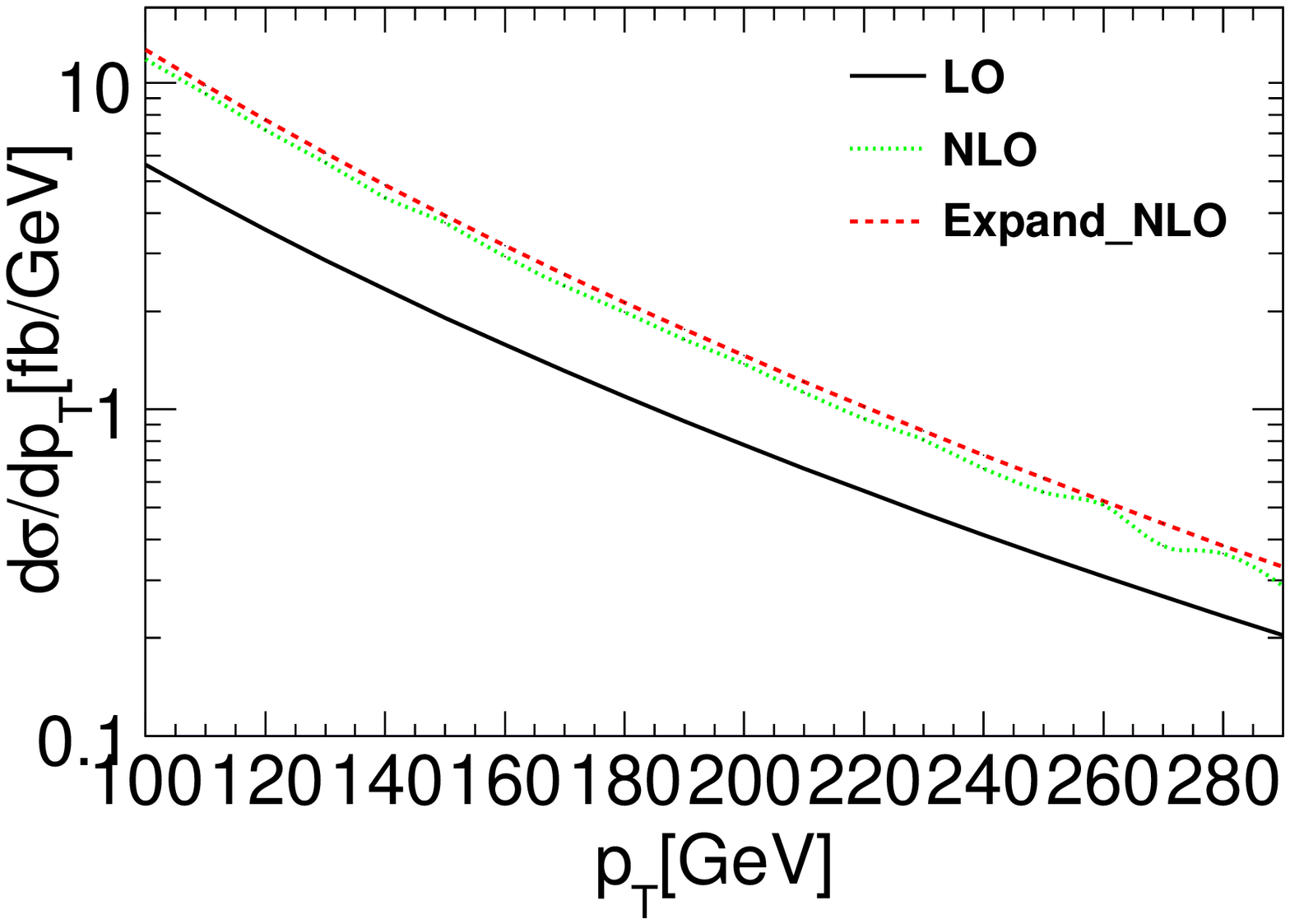}
  \includegraphics[width=0.49\linewidth]{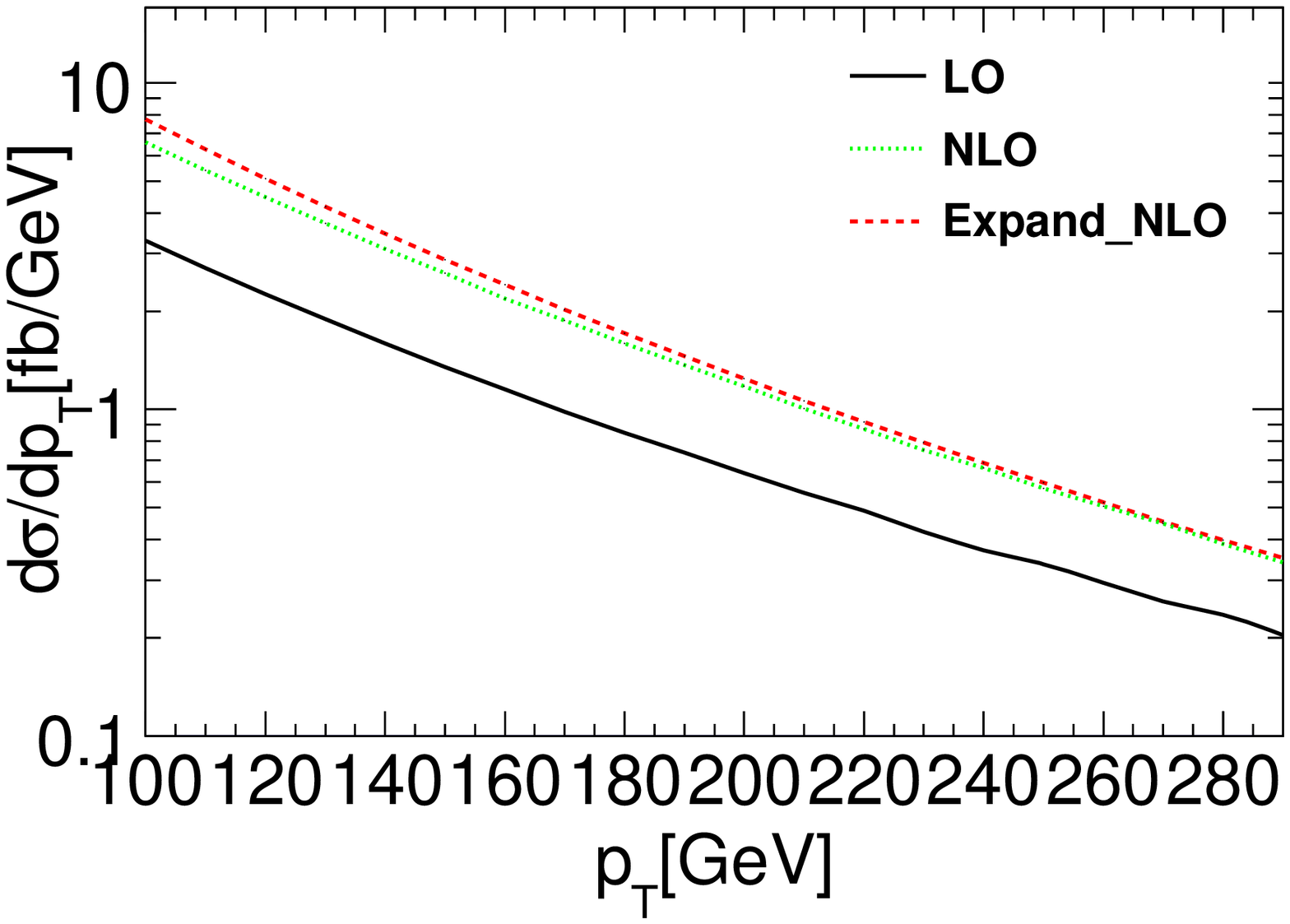}
  \caption{The singular terms and fixed-order  contribution  of the gg channel(left) and gq channel (right) for the Higgs boson and a jet associated production with large $p_T$ at the $8$
  TeV LHC.}
  \label{ggapprox}
\end{figure}
\begin{figure}
  \includegraphics[width=0.8\linewidth]{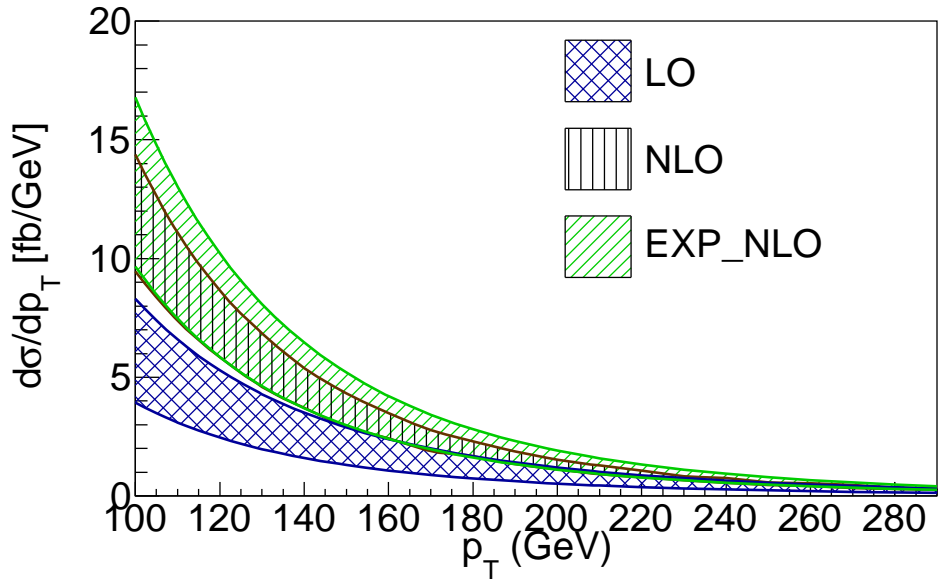}
  \caption{The scale uncertainties of the LO, NLO, expanded NLO results varying the scale from $M_H/2$ to $2 M_H$ for $gg$ channel at the $8$
  TeV LHC.}
  \label{ggexpandscale}
\end{figure}

Before discussing how to choose the different scales for obtaining the numerical
RG-improved cross sections, it is necessary  to
examine to what extent the singular terms approximate the fixed-order calculations.
In Fig. \ref{ggapprox}, we compare the contribution of the singular terms by expanding the resummation formalism with the LO and NLO results at the 8 TeV LHC.
First, we find that the high order corrections for both the  $gg$ and $gq$ channel Higgs boson production
\footnote{For simplicity, we have denoted  the two subprocess of $gq \to H+j $ and $g\bar{q}\to H+j$ as $gq$ channel.} are very large, which means
that higher QCD corrections are important and needed to be included to give a reliable  perturbative prediction.
Second, we see that the NLO cross section is well  approximated by the singular terms when the
$p_T$ of the  Higgs boson is larger than $100$ (200) GeV for the  $gg$ ($gq$) channel.
Since one can not distinguish the quark jet (in $gq$ channel) from the gluon jet (in $gg$ channel),
the two channels should be combined in order to compare with the experimental measurement.
And because the $gg$ channel dominates in the total and differential cross sections, we
will present the resummed prediction for the Higgs boson production in the range $p_T>100$ GeV.
These observations are also true after considering the  scale uncertainties by varying the scale from $M_H/2$ to $2 M_H$, as shown in Fig.~\ref{ggexpandscale}.

\subsection{Scale  choice and matching}

\begin{figure}
  \includegraphics[width=0.9\linewidth]{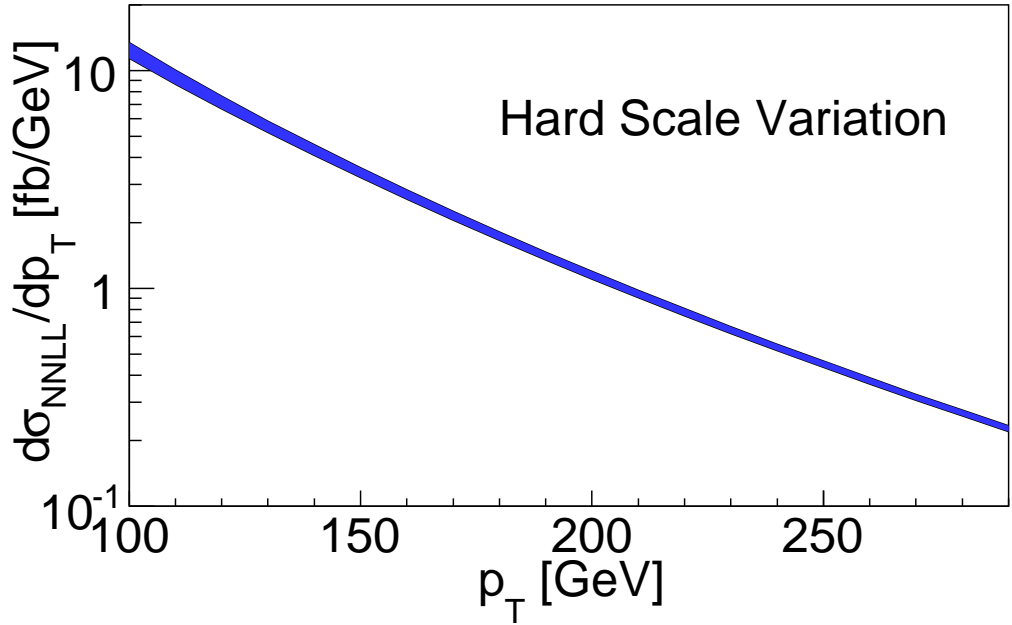}\\
  \caption{The hard scale uncertainty of the resummation results at the $8$ TeV LHC.}
  \label{unhard}
\end{figure}

In the above discussions, the cross section has been factorized into the hard function, jet function and soft function, and each
function only depends on a single scale.
So the hard scale, jet scale and soft scale can be chosen,  respectively, at their intrinsic scales.
Then using the RG, all scales evolve to the same factorization scale.
\begin{figure}
  \includegraphics[width=0.8\linewidth]{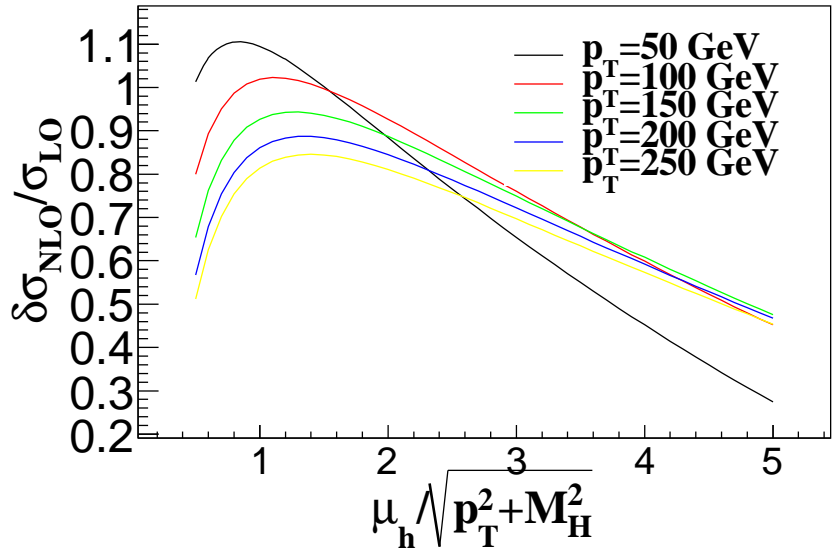}
  \caption{The NLO contribution of the hard function for different $p_T$ cuts at the $8$ TeV LHC.}
  \label{hrdsclrg}
\end{figure}
\begin{figure}
  \includegraphics[width=0.8\linewidth]{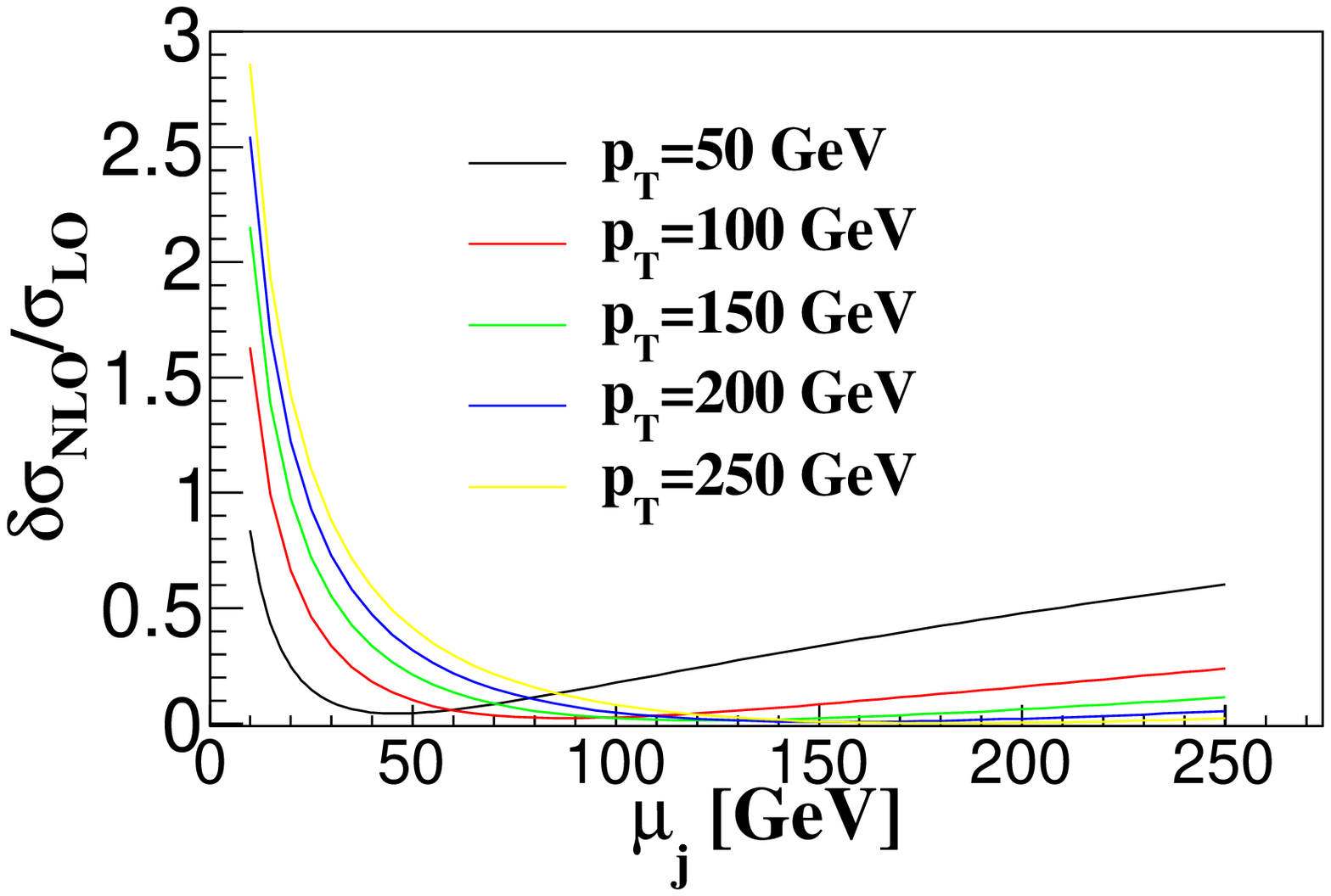}
  \caption{The NLO contribution of the jet function for different $p_T$ cuts at the $8$ TeV LHC.}
  \label{jetsclrg}
\end{figure}
\begin{figure}
  \includegraphics[width=0.8\linewidth]{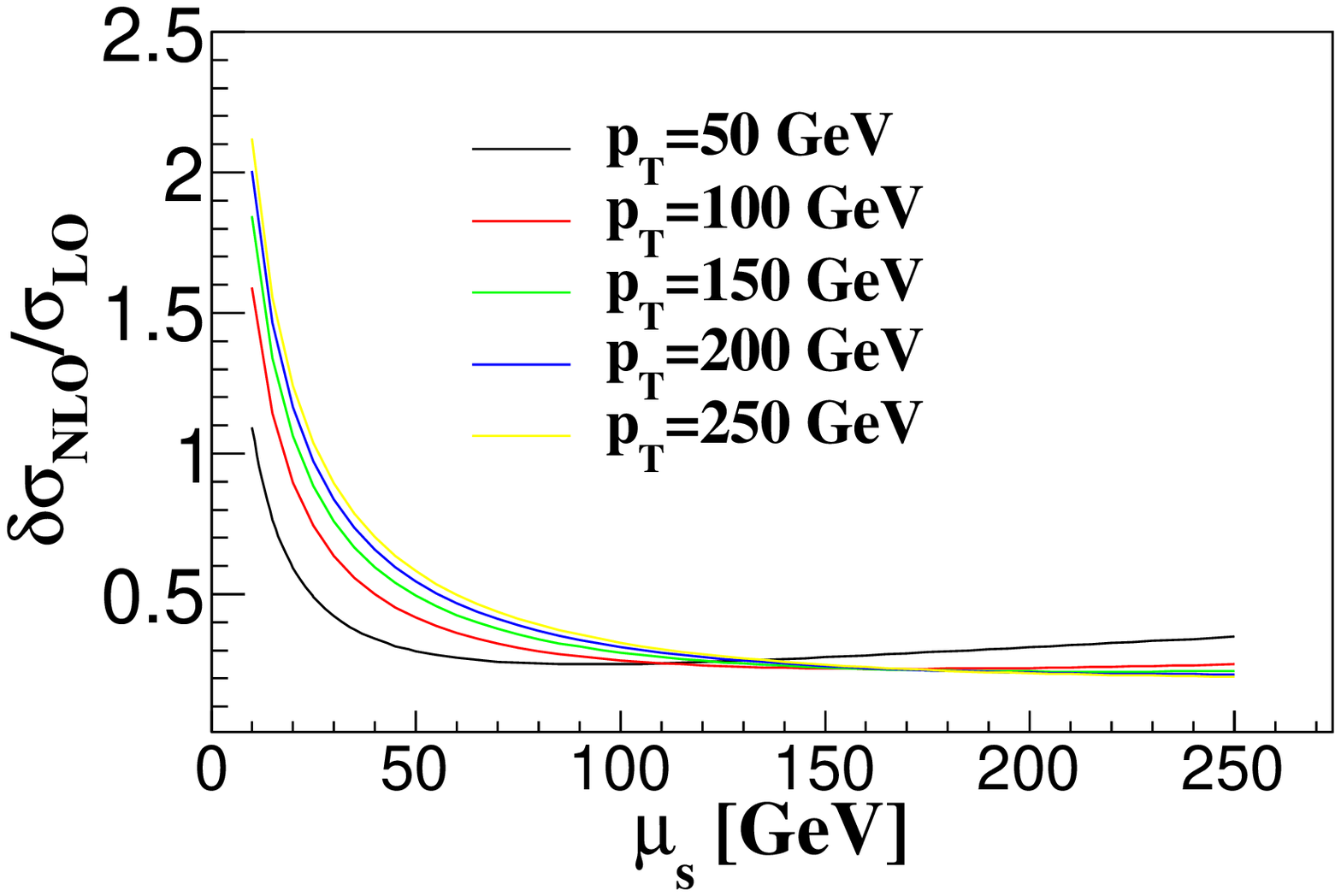}
  \caption{The NLO contribution of the soft function for different $p_T$ cuts at the $8$ TeV LHC.}
  \label{sftsclrg}
\end{figure}
In Fig. \ref{hrdsclrg}, we show the contribution to the NLO correction from only the hard function,
normalized by the LO result, as a function of the hard scale.
The hard function takes maximum values when the hard scale is around $\sqrt{p_{T}^{2}+M_{H}^{2}}$ for $p_T=100 \sim  250$ GeV.
And the contributions from the hard function are very large, generally larger than 0.5 for $p_T =100 \sim  250$ GeV.
This means that the hard function is very important, and needs to be calculated with higher precision.
Resummation is a way to achieve this target.
On the other hand, we also notice that the terms which can be resummed are only a limited part of the hard function.
There is also significant contribution from those terms which are scale independent.
Here, we choose $2.5 \sqrt{p_{T}^{2}+M_{H}^{2}}$ as the default hard scale.
The scale uncertainty of final resummed result from variation of the hard scale is about $12\%$, as shown in Fig.~ \ref{unhard}.

\begin{figure}
  \includegraphics[width=0.9\linewidth]{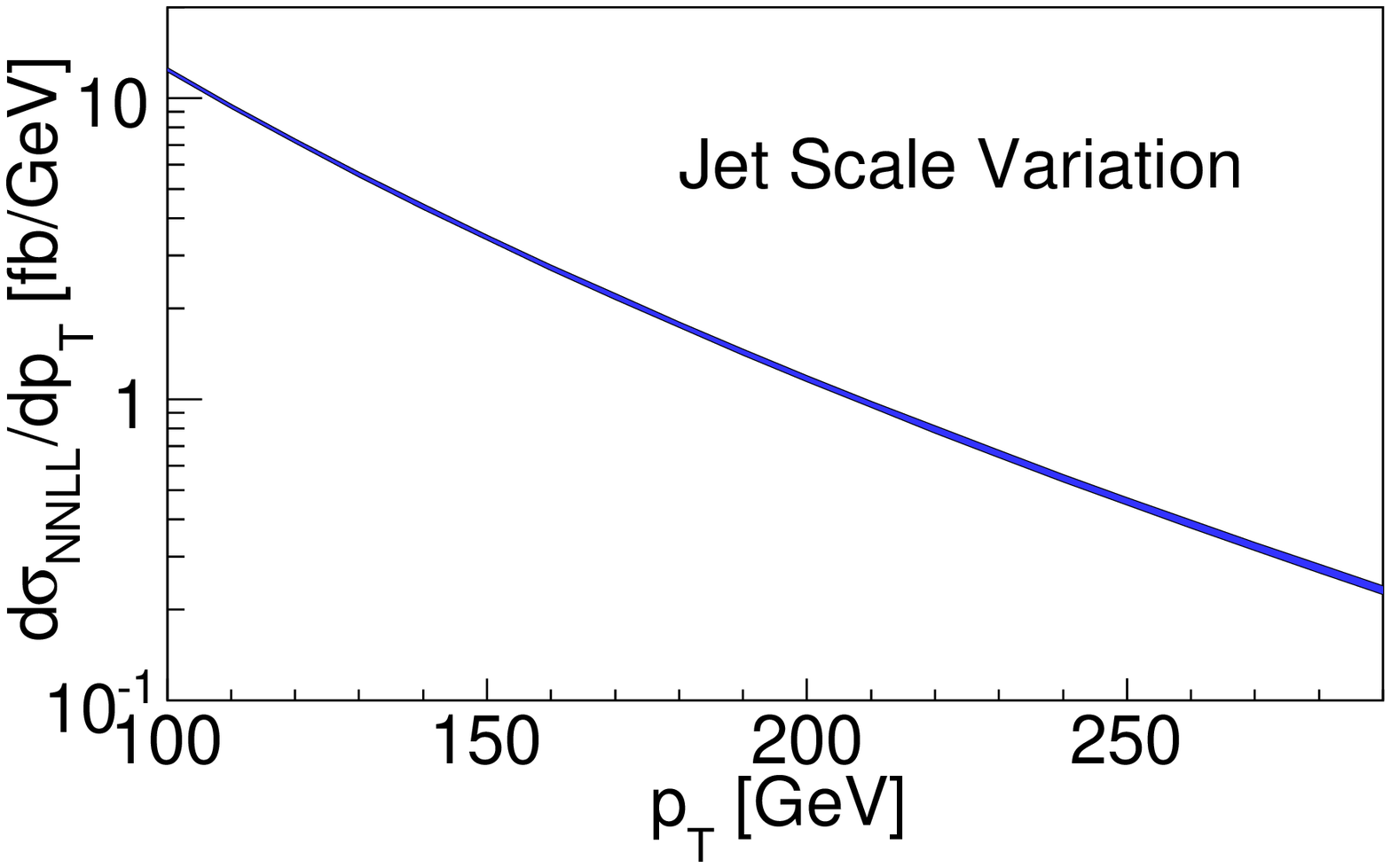}\\
  \caption{The jet scale uncertainty of the resummation results at the $8$ TeV LHC.}
  \label{unjet}
\end{figure}

\begin{figure}
  \includegraphics[width=0.9\linewidth]{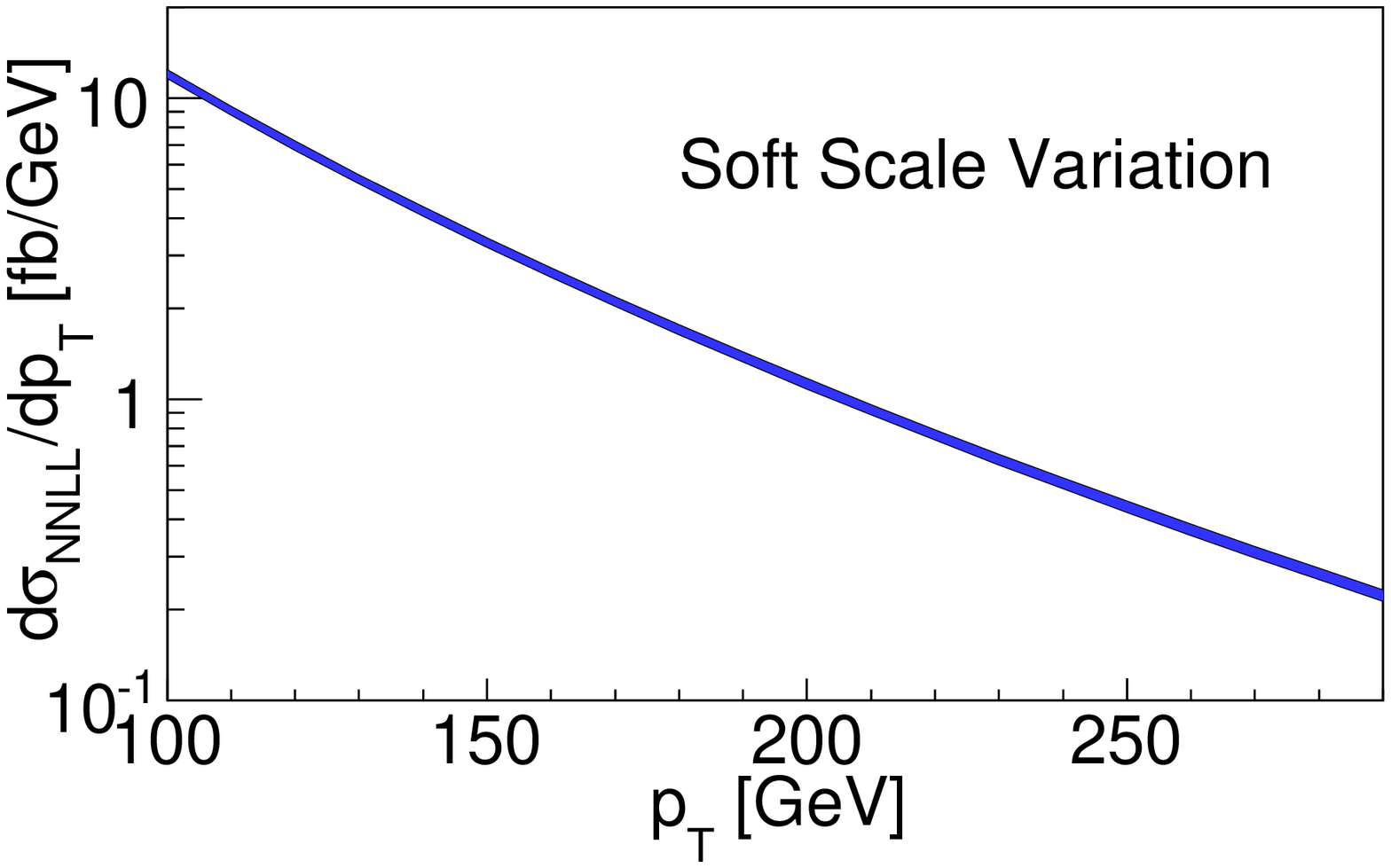}\\
  \caption{The soft scale uncertainty of the resummation results at the $8$ TeV LHC.}
  \label{unsoft}
\end{figure}

In Fig. \ref{jetsclrg}, we show the contribution to the NLO correction from only the jet function,
normalized by the LO result, as a function of the jet scale.
The jet function drops very quickly when the jet scale is smaller than 50 GeV, and changes very slowly when the jet scale is larger than 50 GeV.
We can also see that the contribution from the jet function is  about $10\%$ if $p_T$ is larger than 100 GeV.
The soft function has a similar behavior,  as shown in Fig. \ref{sftsclrg}, except that the contribution from the soft function is about $30\%$ if $p_T$ is larger than 100 GeV.
We choose $ \mu_j \approx150$~GeV, $\mu_s \approx100$~GeV as the default jet and soft scales, respectively.
The uncertainties of the final resummed result from the variation of jet and soft scales are  about $2.4\%$ and $5.8\%$ as shown in Fig. \ref{unjet} and Fig. \ref{unsoft}, respectively.

\begin{figure}
\includegraphics[width=0.79\linewidth]{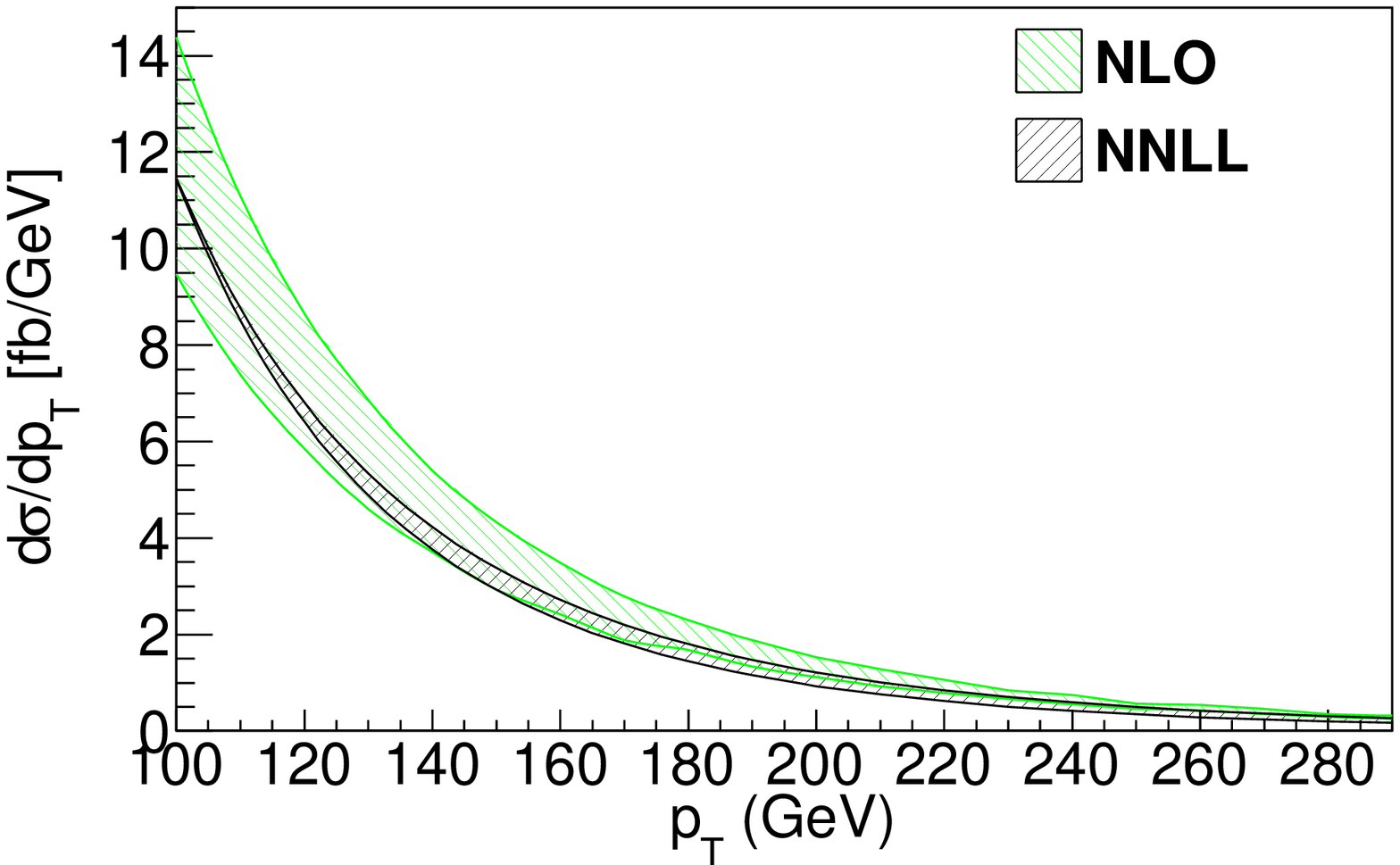}
\caption{The scale uncertainty of the NNLL and NLO  results for the Higgs boson and one jet aasociated production with large $p_T$ at the $8$
TeV LHC for gg channel.}
\label{ggres}
\end{figure}

\begin{figure}
\includegraphics[width=0.79\linewidth]{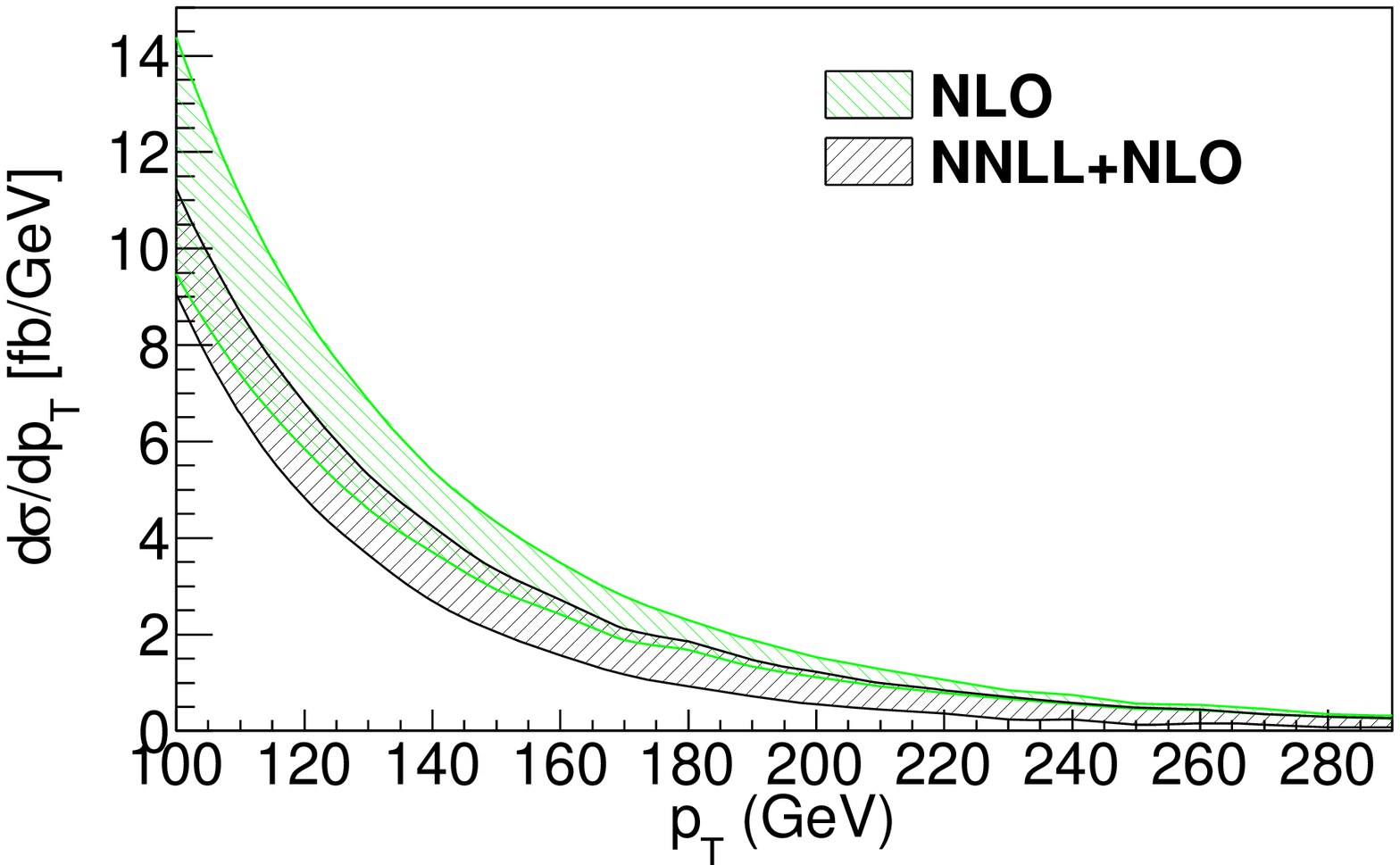}
\caption{The scale uncertainty of the matched and NLO results for the Higgs boson and one jet associated production with large $p_T$ at the $8$
TeV LHC for the gg channel.}
\label{ggmatch}
\end{figure}

Finally,  to examine the factorization scale uncertainty in the final resummed result,
we vary the factorization scale from $M_H/2$ to $2 M_H$ and show the resummed result in Fig.  \ref{ggres}.
For comparation, we present scale uncertainties of the NLO result obtained by varying $\mu=\mu_R=\mu_f=M_H$ by a factor of 2,
as done in Ref. \cite{Boughezal:2013uia}.
After matching the resummed result with the NLO one, as shown in Eq. \ref{eqs:match}, the result is shown in  Fig. \ref{ggmatch}.
We see that both the resummed and matched results have smaller scale uncertainties than the NLO result.


\begin{figure}
  \includegraphics[width=0.8\linewidth]{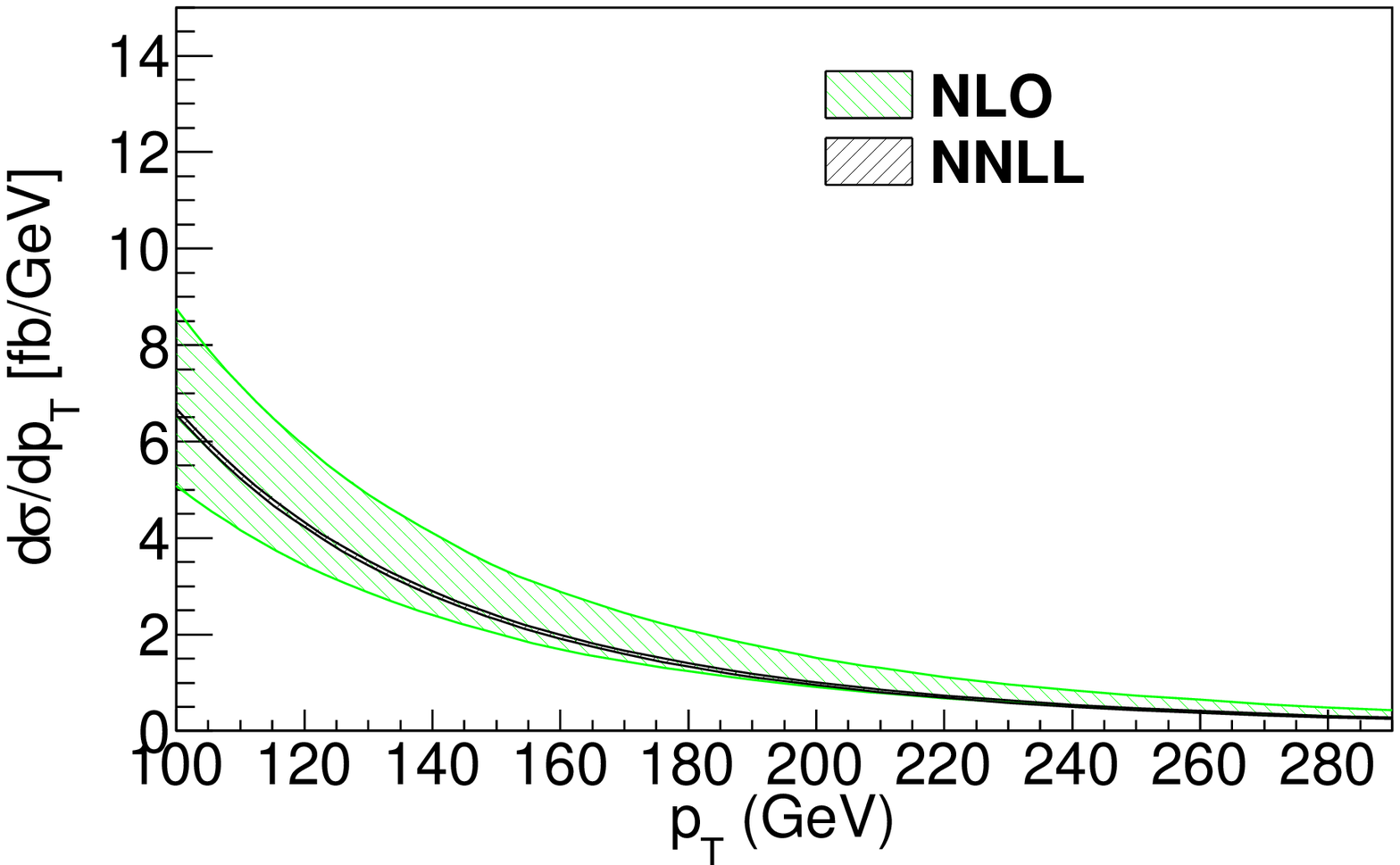}
  \caption{The scale uncertainty of the NNLL and NLO  results for the Higgs boson and one jet associated production with large $p_T$ at the $8$
  TeV LHC for gq channel.}
  \label{gqres}
\end{figure}

\begin{figure}
  \includegraphics[width=0.8\linewidth]{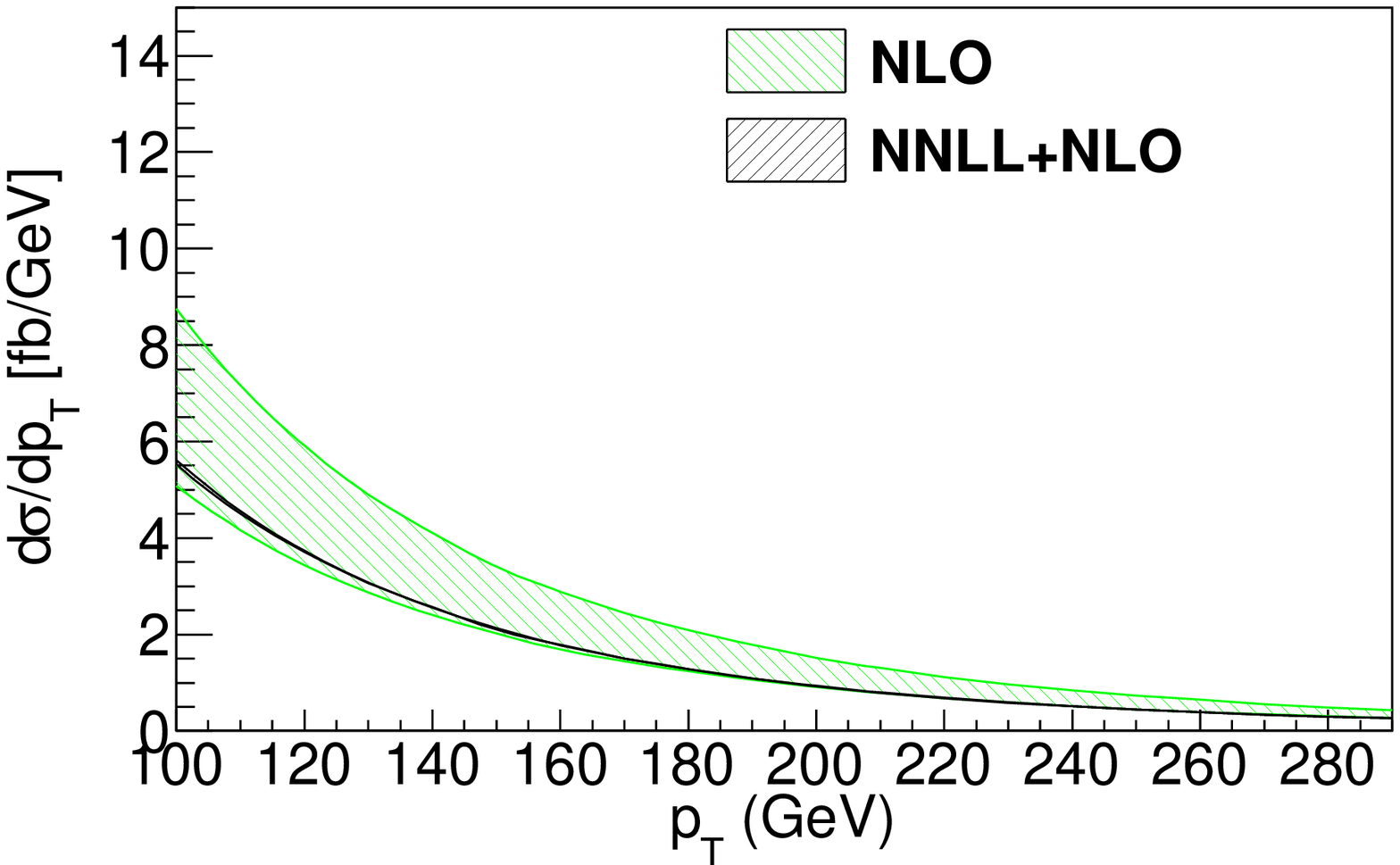}
  \caption{The scale uncertainty of the matched and NLO results for the Higgs boson and one jet associated production with large $p_T$ at the $8$
  TeV LHC for the gq channel.}
  \label{gqmatch}
\end{figure}

The case of $gq$ channel is similar to the $gg$ channel, and the scale uncertainty after resummation reduces more significantly compared with the $gg$ channel;
see Figs. \ref{gqres},\ref{gqmatch}.
For the $q\bar{q}$ channel, the contribution is very small. So we
do not show  its result  individually.
\begin{figure}
  \includegraphics[width=0.8\linewidth]{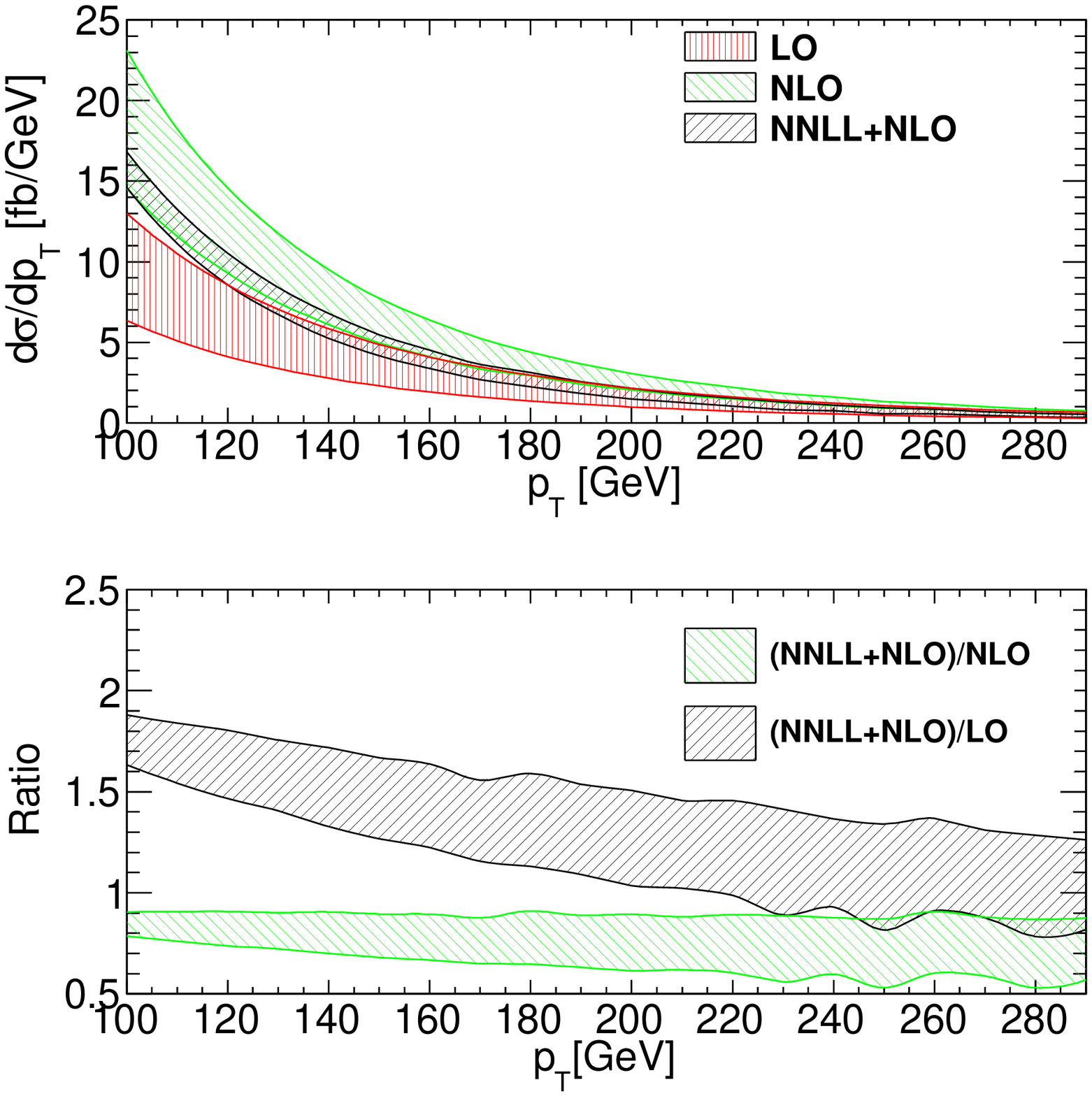}
  \caption{The total scale uncertainty of the matched and NLO results for the single Higgs production with large $p_T$ at the $8$
  TeV LHC.}
  \label{totamatch}
\end{figure}
\newpage

Figure {\ref{totamatch}} shows  the differential cross section as a function of $p_T$ after matching to the NLO results.
The  result with resummation effects obviously reduces the large scale uncertainty compared to the NLO result.
The ratio of the resummation result to the LO cross section is sensitive to $p_T$, changing from 1.8 to 1.0 as $p_T$ varies from 100 to 290 GeV.
The resummation results of the total cross section decrease the NLO one by about $11\%$ at the default scales
when the Higgs boson $p_T$ is larger than 100 GeV.

\subsection{Discussion on the finite top quark mass effects}
\begin{figure}
\includegraphics[width=0.8\linewidth]{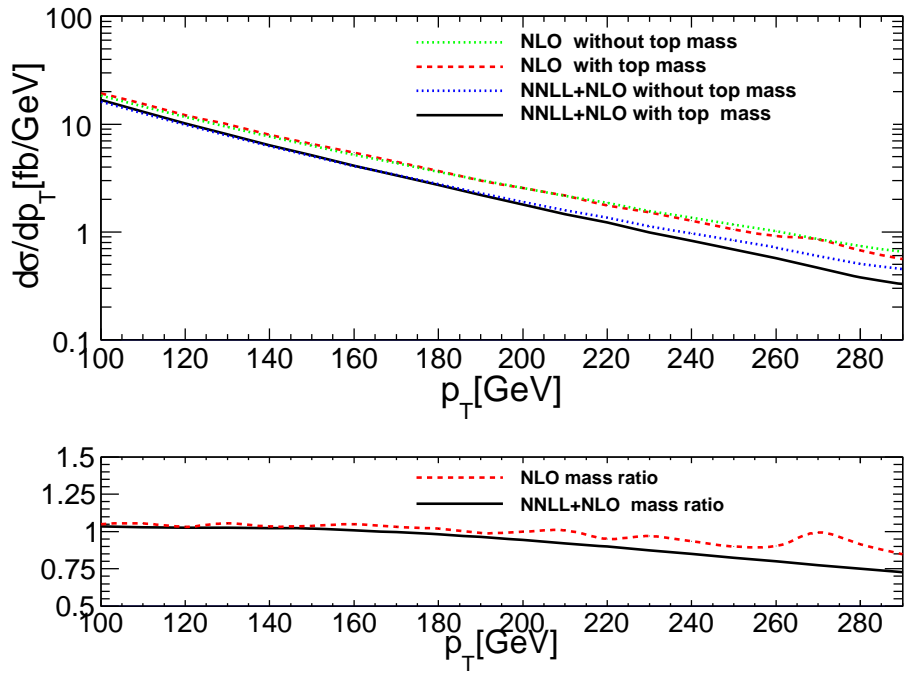}
\caption{The central value ($\mu_F=M_H,\mu_h=2.5 \sqrt{p_{T}^{2}+M_{H}^{2}},\mu_j=150~\rm{GeV},
\mu_s=100~\rm{GeV}$)of four cases for the large $p_T$ Higgs boson production at the $8$
TeV LHC.The green curve is the NLO $p_T$ distribution in the infinite top quark mass limit, the red curve is the NLO one with top
mass effects, the blue curve is the NNLL$+$NLO one in the infinite top quark mass limit and the black curve is the
NNLL$+$NLO with top quark mass effects.}
  \label{topcompare}
\end{figure}

Up to now, the discussion is under the assumption of infinite top quark mass limit.
However, in the case of large $p_T$ Higgs boson production, the
Higgs low energy theorem \cite{Ellis:1975ap,Shifman:1979eb} may fail to apply, and the
top quark mass effects may make a sense. Considering the finite top quark mass  will bring  more complicated calculations,
but also open a new way to probe the coupling of the Higgs to top quarks.
When including the finite top quark mass, there exists only LO results \cite{Ellis:1987xu,Baur:1989cm}
and the corresponding parton shower effects \cite{Alwall:2011cy,Bagnaschi:2011tu}.  Unfortunately,
there is no  complete QCD NLO calculations of the Higgs plus jet including exact top mass
effects. Only the subleading terms in $1/m_t$ have been calculated at QCD NLO \cite{Harlander:2012hf,Grazzini:2013mca}.
It is found that the infinite top quark mass limit is a very good approximation as long
as $p_T<200$ GeV \cite{Ravindran:2002dc, Harlander:2012hf}.
Other discussions can been seen in Refs. \cite{Dittmaier:2012vm,Stewart:2013faa,Harlander:2012hf}.
We investigate the possible top quark mass effects by using the program HNNLO, presenting the results in Figs.\ref{topcompare}.
For Higgs's $p_T $ less than $200$ GeV, the finite top quark mass effects are not obvious
(the difference between the results with and without finite top quark mass is less than $4\%$).
For the $p_T$ larger than $200$ GeV, the top quark mass begins to make sense  and
it is necessary to consider the finite top quark mass effects.
We define the differential $K$ factor as
\begin{equation}
K(p_T)=\frac{d\sigma_{\infty}^{NLO+NNLL}}{d\sigma_{\infty}^{LO}},
\end{equation}
where $\infty$ refers to the infinite top quark mass limit. Since
the differential $K$ factor depends weakly on the top quark mass
\cite{Spira:1995rr,Harlander:2009mq,Pak:2009dg}, we can obtain
a reliable approximation of higher-order cross section by multiplying
the $K$ factor to the exact top mass dependent LO one  following the methods in Refs.
\cite{Grazzini:2013mca,Grigo:2013xya}, which is given by
\begin{equation}
\frac{d \sigma}{d p_T}=\frac{d \sigma_{m_t}^{LO}}{dp_T} K(p_T).
\end{equation}
Here, $\sigma_{m_t}^{LO}$ means the exact LO cross section with finite top quark mass.
The obtained $p_T$ distribution of the Higgs boson at NNLL$+$NLO with finite top quark mass is shown in
Fig. \ref{topcompare} as the black curve  at the central value of the scales.
In Fig. \ref{topcompare}, we compare the  resummation results with NLO ones
in both cases of infinite  and finite top quark mass.
These results can be used to improve the accuracy in probing the Higgs couplings to top quarks in
the recent works \cite{Azatov:2013xha, Banfi:2013yoa,Grojean:2013nya}.

\subsection{Simple discussions on higher order corrections}

\begin{figure}
  \includegraphics[width=0.8\linewidth]{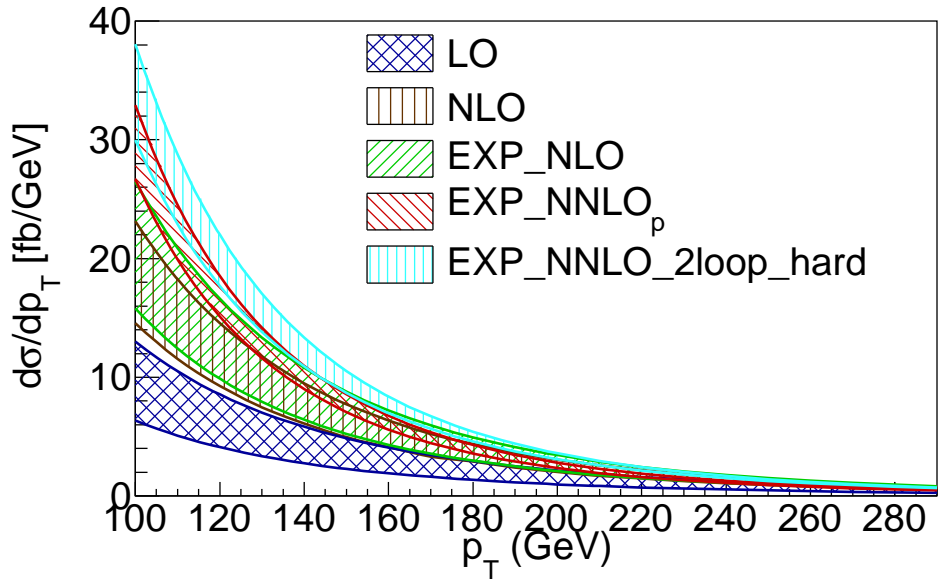}
  \caption{The total scale uncertainties of the LO, NLO, expanded NLO,  expanded $\rm{NNLO_p}$ (with  one-loop hard function squared), and expanded NNLO (with the two-loop hard function ) results varying the scale from $M_H/2$ to $2 M_H$ with all channels at the $8$
  TeV LHC.}
  \label{twoexpand}
\end{figure}

After our paper appeared as an e-print, the authors in Ref. \cite{Becher:2014tsa}
investigated the same process and included the two-loop hard function. Their NNNLL results are not
matched to the NNLO fixed order results in their paper.
We use SCET to resum the large logarithm and match to the
NLO fixed order one,  while the authors in Ref. \cite{Becher:2014tsa} use SCET to predict the approximated NNLO result,
and their main conclusion is that the approximated NNLO correction  increases the NLO by $50\%$.
The  expanded $\rm{NNLO_p}$ results  with
the one-loop hard function squared (NNLO singular terms expanded from one-loop hard function, two-loop jet function, and two loop soft function)
is shown  in Fig. \ref{twoexpand},
and we see that the  expanded $\rm{NNLO_p}$  result with  the one loop hard function squared  increases the NLO one significantly.
The reason is that the main contribution to the two-loop hard function is from the squared one-loop hard function, shown as $A_0^2$ term in $B_0$ from Eq.(\ref{b0}).
Besides, our results of LO and NLO singular terms are exactly the same as their results
if we choose the same parameters as in \cite{Becher:2014tsa}.

Furthermore, we incorporate the two-loop hard function extracted from Refs.\cite{Gehrmann:2011aa,Becher:2013vva},
and find that the expanded NNLO results with two-loop hard function (NNLO singular terms from two-loop hard function,
two-loop jet function and two-loop soft function) are numerically identical to the ones in Ref. \cite{Becher:2014tsa}
when  the same parameters are chosen. The expanded NNLO result with two-loop hard function is shown in Fig. \ref{twoexpand}, where the term $c_2^H-{c_1^H}^2/2$ is added in order to compare with the expanded $\rm{NNLO_p}$ result with  one-loop hard function squared.
Figure \ref{twoexpand} shows that the difference between the two kinds of expanded NNLO results is not very large. This is because
the dominant contribution comes from the large logarithm terms which are obtained by expansion of the RG evolution expression,
not the $c_2^H-{c_1^H}^2/2$.
At NNLL order with the default scales, the contribution of negative logarithm terms in $A_0$  dominates, and this leads to the
fact that the resummation effects decrease the NLO cross section.
After expanding the cross section to NNLO order, new positive logarithms, such as $A_0^2$ terms in $B_0$, overwhelm the negative ones.
Thus, the corrections for the two kinds of expanded NNLO results with the one-loop and two-loop hard function, respectively,
become positive  as shown in Fig.\ref{twoexpand}, whether the $c_2^H-{c_1^H}^2/2$ term
is included or not.
The NNNLL resummation effect matched to
NNLO fixed order deserves to be studied further, but is  beyond the scope of this paper, and left for future work.

\section{Conclusion}
\label{sec:conc}
We have studied the Higgs boson  production at large  $p_T $  at the LHC, including the resummation effects in SCET.
We find that the  resummation effects decrease the NLO cross sections by about $11\%$ at the central values of the scales
when the Higgs boson $p_T$ is larger than $100$~GeV,
and also reduce the scale uncertainty  obviously.
Moreover, we discuss
the top mass effects numerically, and find that the top quark mass effects increase with the increasing of  $p_T$.
The $p_T$ distribution of Higgs boson is  important for describing the Higgs boson production at the LHC, and it is
sensitive to the QCD higher order corrections.
A precise measurement of the Higgs $p_T$ is expected to
be given in the near future
and its precise prediction  is very important for the experimental analyses.
Thus, it is necessary to precisely investigate the large $p_T$ behavior of the Higgs boson, and
any deviation of the Higgs boson's $p_T$ distribution will give hints to the possible modification of the Higgs couplings to  the top quark, which will shed light on the new physics.

\acknowledgments
This work was partially supported by the National Natural
Science Foundation of China, under Grants No. 11375013 and No. 11135003.
Jian Wang was supported by the Cluster of Excellence
{\it Precision Physics, Fundamental Interactions and Structure of Matter} (PRISMA-EXC 1098).

\bibliography{singleHiggs}{}

\end{document}